\documentclass[prd,twocolumn,amsmath,amssymb,floatfix,superscriptaddress,nofootinbib,preprintnumbers]{revtex4-1}

\usepackage{graphicx}
\usepackage{amssymb}
\usepackage{amsmath}
\usepackage{bm}
\usepackage{color}
\usepackage[utf8]{inputenc}
\usepackage{comment}

\DeclareMathAlphabet\mathbfcal{OMS}{cmsy}{b}{n}
\definecolor{darkgreen}{RGB}{50,150,0}
\definecolor{purple}{cmyk}{0.5,0.75,0,0}

\newcommand{\Lr}{\textrm{L}} 
\newcommand{\lnM}{{\ln\!M}}

\newcommand{\refeq}[1]{Eq.~(\ref{eq:#1})}
\newcommand{\refeqs}[2]{Eqs.~(\ref{eq:#1})--(\ref{eq:#2})}
\newcommand{\refEq}[1]{Eq.~(\ref{eq:#1})}
\newcommand{\refEqs}[2]{Eqs.~(\ref{eq:#1})--(\ref{eq:#2})}

\newcommand{\reffig}[1]{Fig.~\ref{fig:#1}}
\newcommand{\refFig}[1]{Fig.~\ref{fig:#1}}
\newcommand{\reffigs}[2]{Figs.~\ref{fig:#1}-\ref{fig:#2}}
\newcommand{\reftab}[1]{Tab.~\ref{tab:#1}}
\newcommand{\refsec}[1]{Sec.~\ref{sec:#1}}
\newcommand{\refapp}[1]{Appendix~\ref{app:#1}}

\newcommand{\del}[1]{\delta_{#1}}
\def\be{\begin{equation}}
\def\ee{\end{equation}}
\def\ba#1\ea{\begin{align}#1\end{align}}
\newcommand{\vs}{\nonumber\\}
\newcommand{\kmax}{k_{\rm max}}

\newcommand{\ihMpc}{h\,{\rm Mpc}^{-1}}

\newcommand{\Comment}[1]{{}}

\definecolor{ultramarine}{rgb}{0.07, 0.04, 0.56}
\definecolor{cadmiumgreen}{rgb}{0.0, 0.42, 0.24}
\definecolor{indigo(dye)}{rgb}{0.0, 0.25, 0.42}
\usepackage[linktocpage=true]{hyperref}
\hypersetup{
colorlinks=true,
citecolor=ultramarine,
linkcolor=cadmiumgreen,
urlcolor=indigo(dye),
pdfauthor={},
pdftitle={},
pdfsubject={}
}

\begin{document}

\preprint{YITP-SB-16-35}

\title{Quintessential Scale Dependence from Separate Universe Simulations}

\author{Chi-Ting Chiang}

\affiliation{C.N. Yang Institute for Theoretical Physics, Department of Physics \& Astronomy,
Stony Brook University, Stony Brook, NY 11794}

\author{Yin Li}

\affiliation{Berkeley Center for Cosmological Physics, Department of Physics
and Lawrence Berkeley National Laboratory, University of California, Berkeley, CA 94720}
\affiliation{Kavli Institute for the Physics and Mathematics of the Universe (WPI),
UTIAS, The University of Tokyo, Chiba 277-8583, Japan}

\author{Wayne Hu}

\affiliation{Kavli Institute for Cosmological Physics, Department of Astronomy \& Astrophysics,  Enrico Fermi Institute, University of Chicago, Chicago, IL 60637}

\author{Marilena LoVerde}

\affiliation{C.N. Yang Institute for Theoretical Physics, Department of Physics \& Astronomy,
Stony Brook University, Stony Brook, NY 11794}

\begin{abstract}

By absorbing fluctuations into a local background, separate universe simulations
provide a powerful technique to characterize the response of small-scale observables
to the long-wavelength density fluctuations, for example those of the power spectrum
and halo mass function which lead to the squeezed-limit $n$-point function and halo
bias, respectively. Using quintessence dark energy as the paradigmatic example, we
extend these simulation techniques to cases where non-gravitational forces in other
sectors establish a Jeans scale across which the growth of density fluctuations
becomes scale dependent. By characterizing the separate universes with matching
background expansion histories, we show that the power spectrum and mass function
responses depend on whether the long-wavelength mode is above or below the Jeans
scale. Correspondingly, the squeezed bispectrum and halo bias also become scale
dependent. Models of bias that are effectively local in the density field at a
single epoch, initial or observed, cannot describe this effect which highlights
the importance of temporal nonlocality in structure formation. Validated by these
quintessence tests, our techniques are applicable to a wide range of models where
the complex dynamics of additional fields affect the clustering of matter in the
linear regime and it would otherwise be difficult to simulate their impact in the
nonlinear regime.

\end{abstract}

\maketitle

\section{Introduction}
The coupling between density fluctuations of different wavelengths is one of the
most important topics in the study of large-scale structure \cite{Bernardeau:2001qr}.
These couplings can be imprinted in the inflationary initial conditions or develop 
through  gravitational evolution.  In the latter class, a long-wavelength density mode
affects the evolution of all short-wavelength modes that are embedded in it leading
to changes in the power spectrum \cite{Takada:2013bfn,Li:2014jra,Chiang:2014oga,Chiang:2015eza}
and the dark matter halo abundance which gives rise to halo bias \cite{Mo:1995cs,Seljak:2012tp}.
In the limit of a large separation in these scales, one can use the ``separate universe''
(SU) approach to describe these and other effects through a change in the background
density within which small scale structure evolves 
\cite{1993MNRAS.262..717B,Baldauf:2011bh,Sherwin:2012nh}.

The SU approach is not only conceptually straightforward to understand but can also
be readily implemented in cosmological simulations, arbitrarily deep into the nonlinear
regime where perturbation theory breaks down \cite{Sirko:2005uz,Gnedin:2011kj,Li:2014sga,Wagner:2014aka}.
In particular, SU simulations have enabled studies of the squeezed-limit $n$-point
correlation functions \cite{Wagner:2015gva} and their impact on the power spectrum
covariance \cite{Li:2014sga}, the halo bias \cite{Li:2015jsz,Lazeyras:2015lgp,Baldauf:2015vio},
and the Lyman-$\alpha$ forest \cite{McDonald:2001fe,Cieplak:2015kra}. Since the whole
time evolution of the long-wavelength mode is properly captured, as opposed to just
a single epoch such as the time of observation, temporally nonlocal effects on small-scale
observables such as the nonlinear power spectrum \cite{Ma:2006zk} and halo bias
\cite{Senatore:2014eva,LoVerde:2014pxa} are correctly modeled.

Previous studies have focused on SU simulations in the $\Lambda$CDM cosmology,
where only matter clusters at low redshift. If the system contains additional
clustering components such as dynamic dark energy or massive neutrinos, then
one has to be careful when applying the separate universe principle. Specifically,
the separate universe construction is only strictly true if long-wavelength
perturbations evolve under gravitational forces alone and not internal stress
gradients \cite{Dai:2015jaa,Hu:2016ssz}. This means that a SU description would
seem to require that the long-wavelength mode be larger than the Jeans or free
streaming scale of the system. On the other hand, if the impact on small-scale
structure of these extra components is only gravitational, then it can be correctly
modeled by matching the local expansion rate to an SU Hubble rate in a ``fake''
SU approach which implicitly requires fictitious energy density components \cite{Hu:2016ssz}.

In this work, we implement and test this multi-component SU method in simulations
with quintessence dark energy. In particular, the growth of long-wavelength matter
fluctuations above or below the Jeans scale of quintessence differs due to clustering
of the dark energy. As a result, the SU expansion history depends on the scale of
the long-wavelength matter fluctuation as does the response of small-scale observables
such as the power spectrum and halo mass function. The latter implies that halo bias
itself will become scale dependent.

The rest of the paper is organized as follows.
In \refsec{theory}, we describe the mapping of perturbations in the quintessence model
onto the SU background above and below the Jeans scale.
In \refsec{sims}, we implement  the SU approach in quintessence simulations.
We present the results of SU simulations in \refsec{pk} and \refsec{bias} for 
the power spectrum response and the  halo bias respectively.
We discuss these results in \refsec{discussion}.
In \refapp{bias_model}, we compare our results to the predictions of scale-dependent halo bias models
in the recent literature.

Throughout the paper, we  adopt a spatially flat cosmology with a Hubble constant
$h=0.7$, matter density $\Omega_m=0.3$, quintessence energy density $\Omega_Q=0.7$,
quintessence equation of state $w_Q=-0.5$, and an initial curvature power spectrum
with scale-invariant tilt $n_s=1$ and amplitude which sets $\sigma_8=1$ today.
These parameters are chosen to highlight the scale dependence of quintessence
rather than for observational viability.

\section{Quintessential Separate Universe}
\label{sec:theory}

Following Ref.~\cite{Hu:2016ssz}, we review here the construction of the separate universe
for the case where components other than the cold dark matter possess Jeans scales. In
\refsec{expansion}, we show that the influence of these components is captured by a modified
expansion history that is defined by the growth history of the large-scale matter density
fluctuation. We apply this construction to quintessence dark energy models in \refsec{quintessence}.

\subsection{Expansion History}
\label{sec:expansion}

A observer sitting within a long-wavelength matter fluctuation $\delta_m$
would measure the {\it local} mean matter density as
\be
 \bar{\rho}_{mW}(a)=\bar{\rho}_m(a)[1+\delta_m(a)] \,,
 \label{eq:rhoW}
\ee
where $W$ denotes a windowed average across a scale much smaller than the
long-wavelength mode. In the SU picture, the local mean evolves as if the
observer were in a SU whose scale factor 
\be
 a_W=\frac{a}{(1+\delta_m)^{1/3}}\approx a\left(1-\frac{\delta_m}{3}\right) \,,
 \label{eq:aW}
\ee
so that $\bar{\rho}_{mW} \propto a_W^{-3}$.
Note that at early times
\be
 \lim_{t\to 0}\delta_m\to 0 \,, ~~ \lim_{t\to 0}a_W\to a \,,
\ee
and the physical conditions of the local and global cosmology coincide. We
have implicitly assumed that there is a universal time coordinate between
the two and so in the relativistic limit $\delta_m$ is specifically the
synchronous gauge density perturbation \cite{Hu:2016ssz}.

Notice that the SU construction requires only $\delta_m(a)$ itself, not the
evolution of any other density component in the universe. The other components 
determine the evolution of $\delta_m(a)$, but they do not enter into $a_W$
explicitly. If these components only influence small-scale observables through
their impact on $\delta_m(a)$, their effects can be characterized by $a_W$
and the {\it local} Hubble expansion
\be
 H_W=\frac{\dot{a}_W}{a_W}=H-\frac13\dot{\delta}_m=H\left(1-\frac13\delta'_m\right) \,,
\label{eq:H_W}
\ee
where $'\equiv d/d\ln a$. This expansion history does not even need to be
given by a SU Friedmann equation involving the local energy densities and
curvature \cite{Gnedin:2011kj,Hu:2016ssz}. With $a_W$ and $H_W$ alone, we
can model the small-scale observables using $N$-body simulations with this
SU expansion rate.

This construction includes cases where the other components experience
non-gravitational forces which define their Jeans scales. In these cases,
the growth history of $\delta_m(a)$ depends on scale. Since the SU expansion
history depends on the whole growth history, regions of different sizes that
share a common $\delta_m$ at a fixed $a$ will produce different responses
in the small-scale observables. In other words, these observables cannot be
described solely by the change in the local density at the time of observation
alone. For example, as we shall see in \refsec{bias}, the response of the
dark matter halo abundance to $\delta_m(a)$ leads to a halo bias that violates
the local bias expectation of scale independence in the linear regime.

\subsection{Quintessence}
\label{sec:quintessence}

Quintessence or scalar field dark energy models provide a simple arena to
explore the response of small-scale observables to long-wavelength fluctuations,
in particular their amplitude, scale, and growth history. The construction
of the SU with quintessence perturbations has been extensively discussed in
Ref.~\cite{Hu:2016ssz}. Here we only summarize the results that are related
to simulating observable responses above and below the quintessence Jeans scale.

The sound speed of quintessence $c_Q$ sets the sound horizon or Jeans scale
$r_J \sim c_Q /a H$ across which its influence on the evolution of $\delta_m$
differs. If $\delta_m$ has a wavelength smaller than $r_J$, the quintessence
perturbation is Jeans stable and becomes negligible in comparison. Thus the
matter fluctuations evolve under
\be
 \del{\downarrow}''+\left(2+\frac{H'}{H}\right)\del{\downarrow}'
 =\frac32\frac{H_0^2}{H^2}\frac{\Omega_m}{a^3}\del{\downarrow}\,,
\label{eq:dm_subJ}
\ee
where $\delta_m=\del{\downarrow}$ and the down arrow in the subscript denotes
the sub-Jeans case. On the other hand, if $\delta_m$ has a wavelength larger
than $r_J$, then the quintessence perturbation $\delta_Q$ has an impact on
$\delta_m$. Assuming that all fluctuations arise from initial curvature
fluctuations and the sound speed of quintessence is much smaller than speed
of light, we have for the two-component system \cite{Hu:2016ssz}
\ba
 \delta'_Q-3w_Q\delta_Q\:&=(1+w_Q)\del{\uparrow}' \,, \vs
 \del{\uparrow}'' +\left(2+\frac{H'}{H}\right)\del{\uparrow}'\:&=
 \frac32\frac{H_0^2}{H^2}\left[\frac{\Omega_m \del{\uparrow}}{a^3}
 +\frac{\Omega_Q \delta_Q}{a^{3(1+w_Q)}}\right] \,,
\label{eq:dm_supJ}
\ea
where $\delta_m =\del{\uparrow}$ and the up arrow  in the subscript denotes
the super-Jeans case. For simplicity we have also taken the quintessence
equation of state parameter $w_Q=\bar p_Q/\bar \rho_Q$ to be a constant.
With the assumed curvature initial conditions, the initial conditions for
the fields are set by taking $\delta_m=\del{\uparrow}=\del{\downarrow}$
and $\delta_Q$ are all proportional to $a$ in the matter dominated limit.

\begin{figure}[h]
\centering
\includegraphics[width=0.45\textwidth]{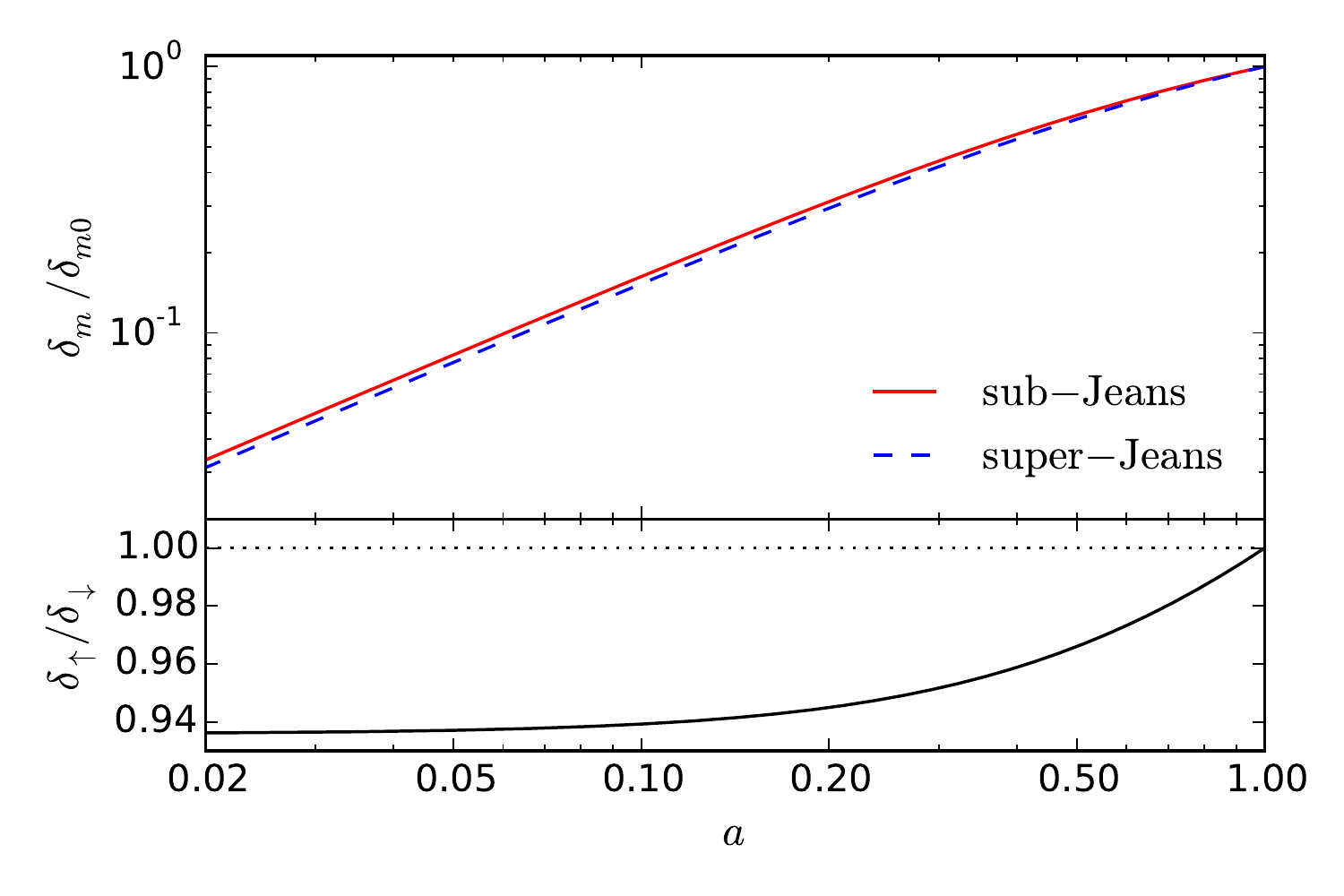}
\caption{(Top) Scale-dependent growth in $\delta_m$ as a function of the scale
factor for super-Jeans (red solid) and sub-Jeans (blue dashed) long-wavelength
modes. (Bottom) The ratio of $\delta_m$ in super-Jeans to sub-Jeans cases. When
characterized as a separate universe, the former is closer to the global universe
than the latter in the past for the same density fluctuation today.}
\label{fig:dm}
\end{figure}

In \refFig{dm} we plot $\delta_m/\delta_{m0}$ as a function of the global
scale factor, where $\delta_{m0}=\delta_m(a=1)$ is the present-day overdensity.
The red solid and blue dashed lines show the sub-Jeans and super-Jeans SUs.
Normalized to the same $\delta_{m0}$, the super-Jeans SU is always closer
to the global universe ($\delta_m=0$) in the past than the sub-Jeans SU in
its expansion history. This implies that the response of the small-scale
observables such as the power spectrum and halo abundance should be smaller
in the super-Jeans than the sub-Jeans SU.

Finally for setting up simulations in the next section it is useful to define
the linear growth of short-wavelength structure in the SU. If the small-scale
matter fluctuations of interest are well within $r_J$, the growth function $D_W$
is simply \refeq{dm_subJ} with the SU expansion history
\be
 \frac{d^2D_W}{d\ln a_W^2}+\left(2+\frac{d\ln H_W}{d\ln a_W}\right)\frac{dD_W}{d\ln a_W}
 =\frac32\frac{H_{0W}^2}{H_W^2}\frac{\Omega_{mW}}{a_W^3}D_W \,.
\label{eq:DW}
\ee
Note that 
\be
 \Omega_{mW} H_{0W}^2 = \Omega_m H_0^2 \,,
\ee
and so only the SU expansion rate $H_W$ from \refeq{H_W} is required to solve
for $D_W(a_W)$. As we shall see next, we can generalize this statement to nonlinear
observables with SU $N$-body simulations.

\section{Separate Universe Simulations}
\label{sec:sims}

The growth history of the long-wavelength matter fluctuation $\delta_m(a)$ in
the global universe alone sets the expansion history in the separate universe.
All effects from long-wavelength fluctuations in other species are incorporated
into its growth history. In the quintessence model, within the Jeans scale dark
energy perturbations can be ignored and so the response of small scale observables
can be calibrated using $N$-body simulations with just this change in the expansion
history.

Unlike the SU technique in $\Lambda$CDM, the change of cosmological parameters 
and their correspondence with real energy densities and curvature becomes non-trivial
for the quintessence model \cite{Hu:2016ssz} whereas the direct change in the
expansion rate $H_W$ remains simply determined by $\delta_m(a)$. Thus, while
some steps are similar to the $\Lambda$CDM SU techniques for running and analyzing
simulations (see e.g. \cite{Li:2014sga,Wagner:2014aka,Li:2015jsz,Lazeyras:2015lgp}),
there are some major differences in performing the SU simulations with quintessence
which we now describe.

Let us start with setting the initial conditions for the simulations. Recall that 
at high redshift the separate and global universes are identical in their physical
description. To achieve this, we first compute the linear power spectrum with the
global cosmology at $z=0$ using CAMB \cite{Lewis:1999bs,Howlett:2012mh}. We then
rescale this power spectrum to the initial redshift of the simulations $a_{Wi}=0.02$ as
\be
 P_W(k,a_{Wi})=P(k,a_0)\left[\frac{D_W(a_{Wi})}{D(a_0)}\right]^2 \,,
\ee
where $D$ is the linear growth in the global universe and $D_W$ is the linear
growth of the SU following \refEq{DW}. The growth functions are normalized in
the matter dominated epoch as
\be
 \lim_{a \rightarrow 0} D(a)=a, \quad \lim_{a_W \rightarrow 0} D(a_W) = a_W \,.
\label{eq:D_ic}
\ee
Note that $D_W$ in sub-Jeans and super-Jeans SUs are different, as they have different
expansion histories. Another subtlety is that the change in the expansion rate of the
SU makes the traditional unit of comoving $[\ihMpc]$ inconvenient. Throughout this paper
we avoid this confusion by using units of comoving $[{\rm Mpc}]$ and convert for code
purposes as necessary. Given the different scale factors $a$ and $a_W$, the correspondence
between comoving wavenumber and physical wavenumber in the global universe differ. Since
this represents a simple dilation of scales, we can account for it in the interpretation
of observable responses rather than in the simulations directly \cite{Li:2014sga}.

The initial conditions are then set up using realizations of Gaussian random
fields for the primordial fluctuations and evolved to $a_{Wi}$ using second-order
Lagrangian perturbation theory (2LPT) \cite{Crocce:2006ve}. Usual 2LPT codes,
such as the publicly available 2LPTIC \cite{2lptic}, compute the linear growth and
growth rate $f_W=d\ln D_W/d\ln a_W$ at $a_{Wi}$ from the cosmological parameters.
We modify the pipeline such that $D_W$ and $f_W$ from the numerical solution of
\refEq{DW} determine the initial positions and velocities of the particles.

We use Gadget-2 \cite{Springel:2005mi} to carry out the simulations. Standard
Gadget-2 computes the Hubble expansion as a function of the scale factor using
the input cosmological parameters. Instead of finding the corresponding cosmological
parameters, we first compute $H_W$ as a function of $a_W$ with \refEq{H_W} and
\refEqs{dm_subJ}{dm_supJ}, pass the table $(a_W,H_W)$ to the code, and then
interpolate the value of $H_W(a_W)$ when necessary\footnote{Specifically, we
only need to modify \texttt{driftfac.c} and \texttt{timestep.c}. Also since
Gadget-2 checks the consistency of the input parameters, we provide the SU
$\Omega_{mW}$ as well as $h_W$, and $L_W$ where $L_W$ is the box size of the
simulations}. We have verified that the SU results are in excellent agreement
with those of the standard 2LPTIC Gadget-2 pipeline in $\Lambda$CDM where the
SU is implemented by varying cosmological parameters.

Following the procedures in Ref.~\cite{Lazeyras:2015lgp}, we identify halos with
the Amiga Halo Finder \cite{Knollmann:2009pb,Gill:2004km}, which is based on the
spherical overdensity algorithm. The key quantity of the spherical overdensity
algorithm is the density threshold, and we set it to be $\Delta=200$ in the global
universe. To match halos identified in the global cosmology, the threshold relative
to the mean in the SU needs to be rescaled as \cite{Li:2015jsz,Lazeyras:2015lgp}
\be
 \Delta_W=\frac{\Delta}{1+\delta_m(t)}\approx\Delta[1-\delta_m(t)] \,.
\ee
In other words, in the overdense (underdense) universe the threshold becomes smaller
(larger) due to the background fluctuations. From each simulation we obtain one halo
catalog, and we consider only halos with more than 400 particles. We also neglect
sub-halos for simplicity.

In this paper, we perform both the sub-Jeans and super-Jeans SU simulations with
$\del{\uparrow\downarrow 0}=\del{\uparrow\downarrow}(a=1)=\pm0.01$, totaling 4
simulations per set. For each of the 20 sets, we fix their initial phases so that
when we take the difference of the observables between overdense and underdense
SU simulations a large amount of noise due to sample variance is removed. We also
run 40 simulations of the global $\delta_m=0$ universe in order to characterize
the clustering bias for comparison in \refsec{clusteringbias}. The first 20 have
the same initial phases as their SU counterparts. For each of these sets we take
a comoving box size $L=1000\,$Mpc and number of particles $N_p=1024^3$, denoted
as small-box.

We also run 20 simulations with $L=2800$ Mpc and $N_p=1024^3$ particles in the global
universe, denoted as big-box simulations. These big-box simulations are used to measure
the position-dependent power spectrum \cite{Chiang:2014oga} for comparison with the
power spectrum response of the sub-Jeans simulations. The details of the simulations
are summarized in \reftab{sims}.

\begin{table}[h]
 \begin{tabular}{c c c c c c c}
  \hline
  type & SU & $L$ [Mpc] & $N_p$ & $\del{m0}$ & $N_{\rm sets}$\\
  \hline
  small-box & $\uparrow\downarrow$  &1000 & $1024^3$ & $\pm0.01$ & 20 \\
  small-box &  no &1000 & $1024^3$ & 0 & 40 \\
  big-box &  no&  2800 & $1024^3$ & 0 & 20 \\
  \hline
 \end{tabular}
 \caption{Summary of the simulations.}
\label{tab:sims}
\end{table}

\section{Power Spectrum Response}
\label{sec:pk}

In this section, we calibrate the responses in the locally measured power spectrum
to a long-wavelength mode above and below the Jeans scale. In \refsec{resp} we extract
these responses from the SU simulations and show that they are scale dependent and
smaller for modes above the Jeans scale than below. We test these responses against
predictions from perturbation theory in \refsec{pt} and the local, position-dependent,
power spectrum from the big-box simulations with long-wavelength sub-Jeans scale modes
in \refsec{ibn}. The good agreement implies that the SU simulation technique provides
accurate predictions for these small scale observables without the need for direct
simulations of quintessence clustering.

\subsection{Separate Universe Calibration}
\label{sec:resp}

In the presence of a long-wavelength density fluctuation $\delta_m$, the power spectrum
observed locally will differ from the global average. We can characterize the fractional
change in the local power spectrum  as a ``response'' $R_{\rm tot}$ to  $\delta_m$
\be
 \frac{\Delta P}{P} \approx \frac{d \ln P}{d\delta_m} \delta_m \equiv R_{\rm tot} \delta_m \,.
\ee
Since to the leading order $R_{\rm tot}$ is independent of $\delta_m$, it can be calibrated
using the SU simulations once and for all rather than with simulations that follow the
dynamics of individual long-wavelength modes. This is especially advantageous for quintessence,
where super-Jeans modes require simulations with quintessence clustering.

This effect can be observed in a local sample of our universe by dividing it into subvolumes
and measuring the correlation between the local power spectra and the subvolume mean overdensities,
which is known as the position-dependent power spectrum \cite{Chiang:2014oga,Chiang:2015eza}.
Even if only the undivided volume is employed, the coherent change in the local power spectrum
$\Delta P(k)$ due to wavelengths larger than the sample induces a ``super-sample'' covariance
between measurements of different $k$ modes \cite{Takada:2013bfn,Li:2014sga,Li:2014jra}.

In practice, the calibration of the total response with SU simulations involves three pieces:
growth, dilation, and reference-density \cite{Li:2014sga}
\be
 R_{\rm tot}  = R_{\rm growth} + R_{\rm dilation} + R_{\bar \rho} \,.
\ee
$R_{\rm growth}$ describes the change in the growth of a small-scale density fluctuation
at a fixed comoving $k$ in the separate and global universe relative to their own scale
factors. $R_{\rm dilation}$ changes the scale to a fixed wavenumber in the global universe
or physical wavenumber in each. Finally $R_{\bar \rho}$ accounts for the different mean
density of the two universes in the definition of the density fluctuation.

To measure the growth response from SU simulations, we first distribute the dark matter
particles onto a $1024^3$ grid by the cloud-in-cell (CIC) density assignment scheme to construct
the density fluctuation, and Fourier transform the density fluctuations with FFTW \cite{fftw}
to form the power spectrum. For each set of super $\uparrow$ or sub $\downarrow$ Jeans
scale SU simulations with the same initial phases, we estimate the growth response,
\be
 R_{\rm growth,\uparrow\downarrow} \equiv 
 R_{\uparrow\downarrow}
 \equiv \frac{d\ln P_{\uparrow\downarrow}}{d\del{\uparrow\downarrow}}
\ee
as
\be
 \hat{R}_{\uparrow\downarrow}(k,a)=
 \frac{\hat{P}_{\uparrow\downarrow}(k,a|{\scriptstyle +}\del{\uparrow\downarrow,0})
 -\hat{P}_{\uparrow\downarrow}(k,a|{\scriptstyle -}\del{\uparrow\downarrow,0})}
 {2\hat{P}(k,a) \del{\uparrow\downarrow}(a)} \,,
\ee
where we difference the overdense and underdense pairs for each $|\del{\uparrow\downarrow,0}|$.
We then compute the variance of $\hat{R}_{\uparrow\downarrow}$ from the 20 small-box realizations.

The dilation response accounts for the fact that the same comoving $k$ in the SU corresponds
to a different physical $k$ in the global universe. Given the change in the scale factor from
\refeq{aW}, it is analytically related to the local slope in power spectrum as \cite{Li:2014sga}
\be
 R_{\rm dilation}(k,a)=-\frac{1}{3}\frac{d\ln k^3P(k,a)}{d\ln k} \,.
\label{eq:Rdilation}
\ee
To compute the dilation response from simulations, we take the log-derivative of the
mean power spectrum measured from 40 small-box simulations with the global cosmology.
As a result, the dilation response is the same in both sub-Jeans and super-Jeans SUs.
Finally the $\bar\rho$ response is due to the change in the definition of a density
fluctuation
\be
 \delta_m \equiv \frac{\delta\rho_m}{\bar\rho_m} = \delta_{mW} \frac{ \bar\rho_{mW}}{ \bar\rho_m} \,,
\ee
so that to the leading order \refeq{rhoW} implies
\be
 R_{\bar \rho}(k,a)=2 \,.
\label{eq:Rreference}
\ee
Note that the last two responses, dilation and reference-density, do not involve
the SU simulations.

\begin{figure}[h]
\centering
\includegraphics[width=0.46\textwidth]{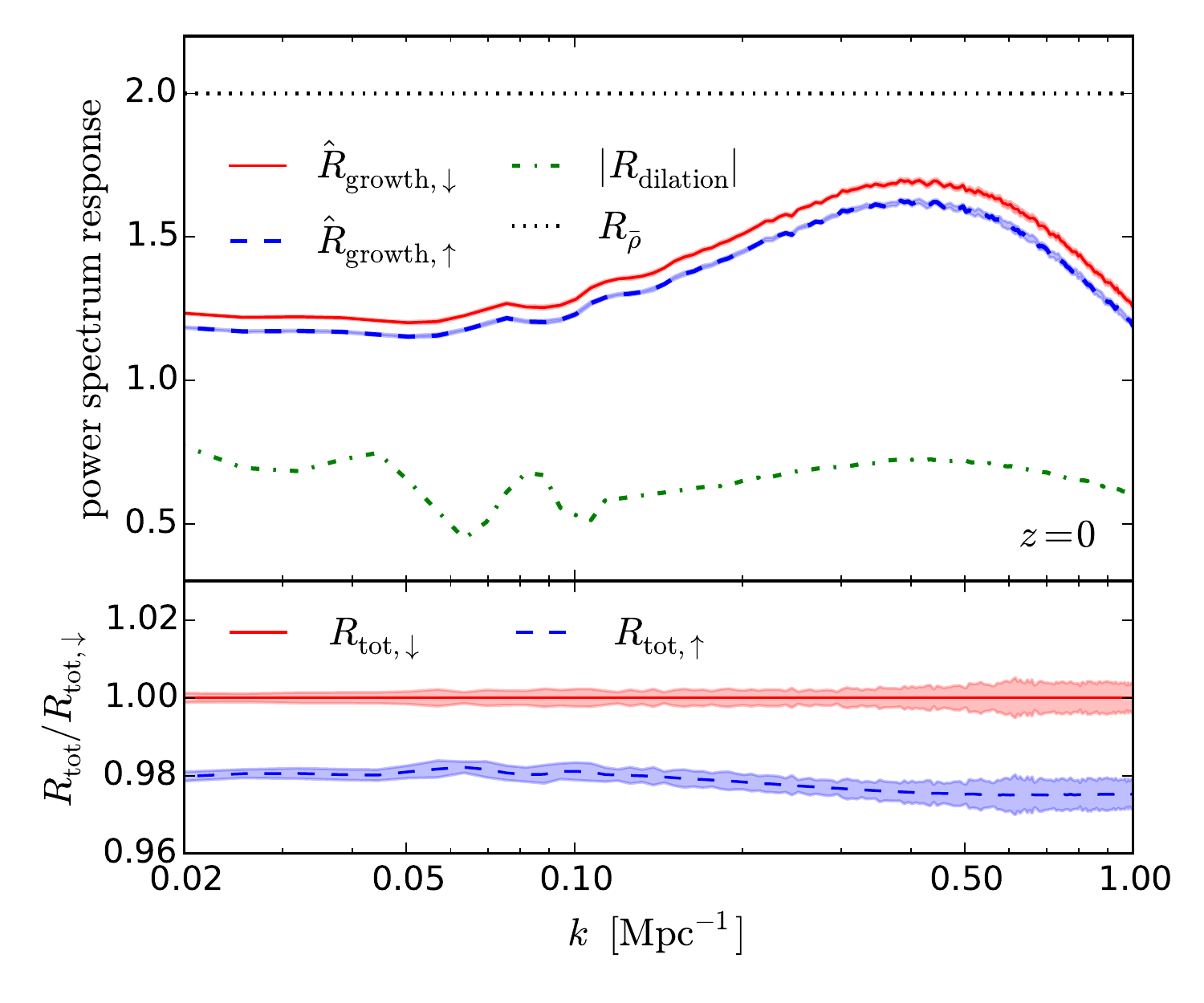}
\caption{(Top) Different components of the separate universe power spectrum responses
at $z=0$: growth response (sub-Jeans, red solid; super-Jeans blue dashed), absolute
value or negative of the dilation response (green dot-dashed), and the reference-density
response (black dotted). (Bottom) Ratios of the mean of the total response to that of
the sub-Jeans separate universe. Shaded bands reflect the error on the mean response
of the simulations. The clear distinction between the sub-Jeans and super-Jeans power
spectrum responses is the first important result of our separate universe simulations.}
\label{fig:resp_tot}
\end{figure}

The top panel of \reffig{resp_tot} shows the various power spectrum responses at $z=0$,
and the bottom panel shows the ratios of the total power spectrum response, $R_{\rm tot}$,
to that of the sub-Jeans SU, $R_{{\rm tot},\downarrow}$. We find that the response is
roughly 2\% smaller in super-Jeans than in sub-Jeans SUs, and the distinction is statistically
significant. Note also the small errors ($\sim0.1\%$ at low-$k$ and $\sim 0.3\%$ at high-$k$)
estimated from small-box SU simulations, demonstrating the power of the SU technique to
precisely characterize the response down to arbitrarily small scales.

The fact that the growth response is smaller in super-Jeans than in sub-Jeans SUs can be
understood qualitatively from \reffig{dm}. Normalized to a given observation redshift,
$\delta_m$ in the super-Jeans limit is always smaller in the past than that in the sub-Jeans
limit. Consequently, the super-Jeans SU is closer to the global universe along the growth
history, and so the response is smaller.

This difference between super-Jeans and sub-Jeans scales produces an observable change
in the local power spectrum, and so can in principle be used as a new probe of the sound
speed of quintessence. In the real universe, the small-scale power spectrum responds
to long modes of all scales, the difference of the responses in sub-Jeans and super-Jeans
limit would thus appear as the scale-dependent squeezed-limit bispectrum for a fixed
small-scale mode. However in quintessence models with initial curvature perturbations,
the predicted amplitude for adiabatic quintessence fluctuations is proportional to
$(1+w_Q)$ (see Eq.~(95) of Ref. \cite{Hu:2016ssz}) and so goes to zero as $w_Q \rightarrow -1$.
More generally, this growth history dependence demonstrates that the nonlinear matter power
spectrum cannot simply be a functional of the linear power spectrum at the same epoch as
is commonly assumed in simple halo model and nonlinear fitting procedures (see also \cite{Ma:2006zk}).

\subsection{Perturbation Theory}
\label{sec:pt}

\begin{figure}[h]
\centering
\includegraphics[width=0.45\textwidth]{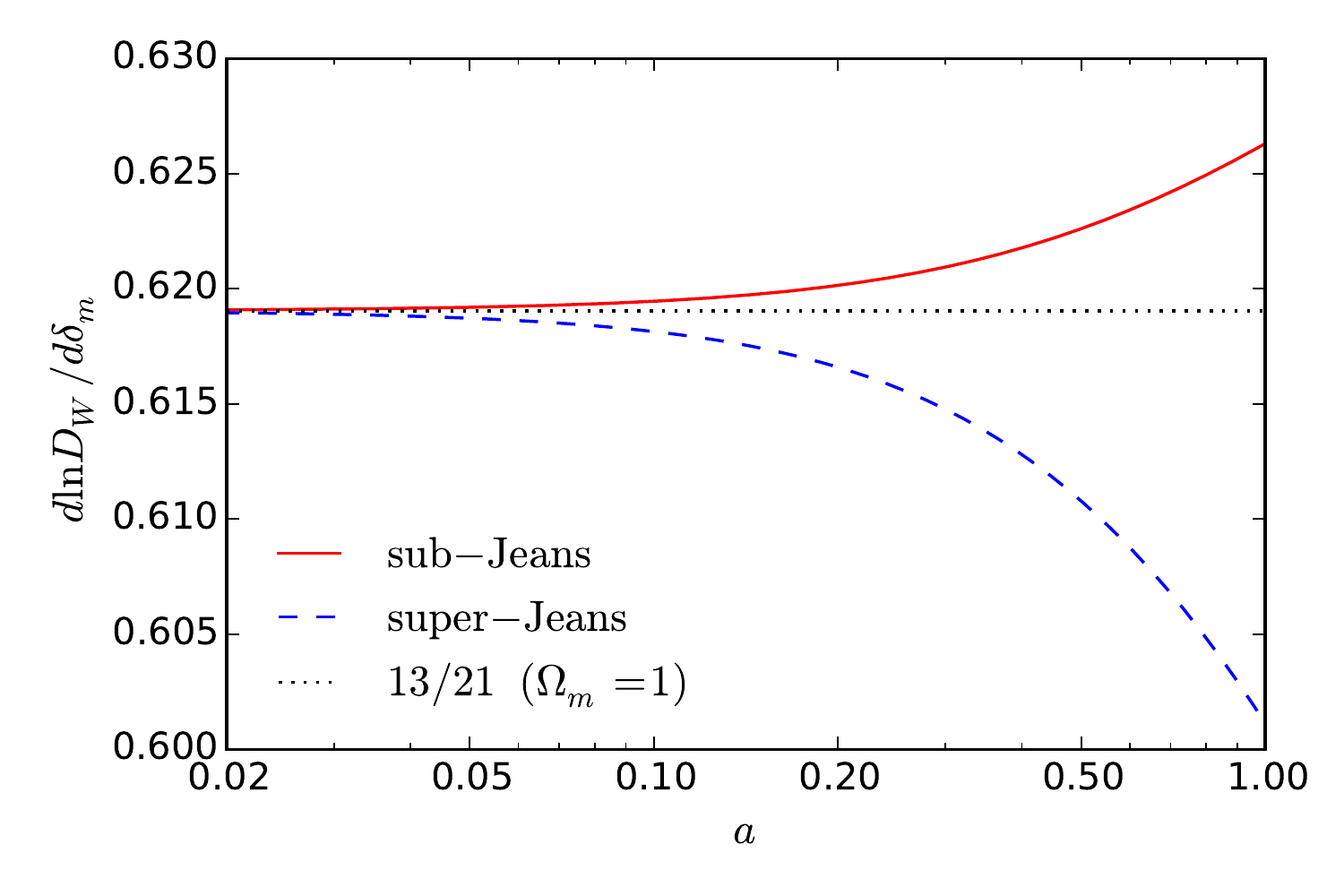}
\caption{The response of the separate universe growth function as a function of the global
scale factor $a$ in the global universe for super-Jeans (red solid) and sub-Jeans (blue
dashed) cases. Above the Jeans scale, the response to the same $\delta_m$ at $a$ is smaller
than below.}
\label{fig:dlnDW}
\end{figure}

\begin{figure}[h]
\centering
\includegraphics[width=0.45\textwidth]{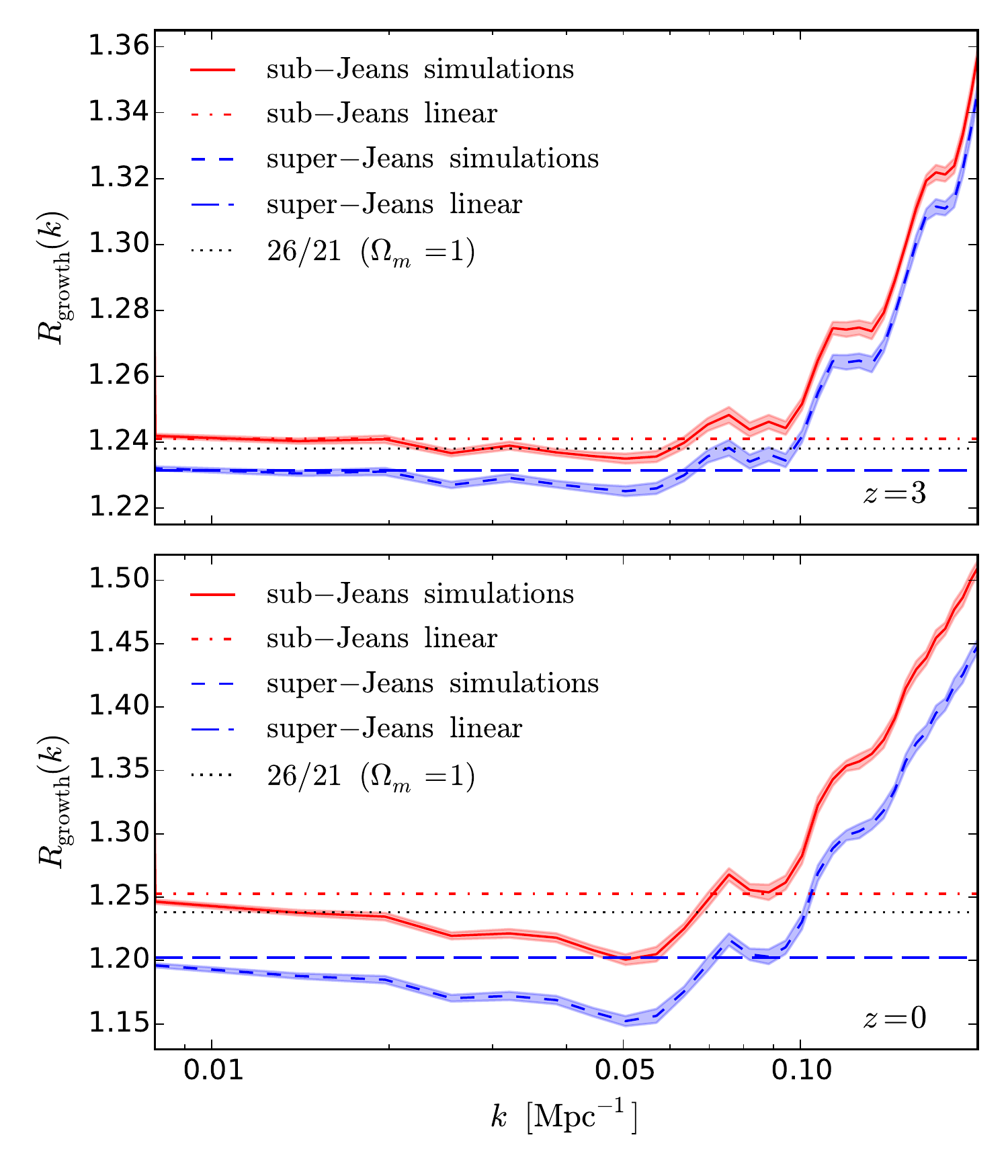}
\caption{Linear perturbation theory predictions for the growth responses compared with
the measurements from the 20 small-box separate universe simulations at $z=3$ (top)
and 0 (bottom). The red solid and blue dashed lines with shaded areas show the sub-Jeans
and super-Jeans separate universes measurements as in Fig.~\ref{fig:resp_tot}, whereas
the red dot-dashed and blue long-dashed lines show the corresponding linear perturbation
theory predictions, i.e.~\refeqs{Rgrowth}{linear}. Note that the range of $y$-axes is
smaller in the top than in the bottom panel. Also the cusp feature at $k\sim0.078~{\rm Mpc}^{-1}$
is a visual artifact due to binning, which we choose to be $\Delta k= 2\pi/L$.}
\label{fig:dlnpk_linear}
\end{figure}

To better understand the growth responses quantitatively, we compute them in perturbation
theory and check their agreement with the SU simulations at various redshifts. In perturbation
theory, the effect can be modeled through the SU linear growth function $D_W$ as
\be
 R_{\rm growth}(k,a)=\frac{d\ln P(k,a)}{d\ln D_W(a)}\frac{d\ln D_W(a)}{d\delta_m(a)} \,.
\label{eq:Rgrowth}
\ee
In the linear regime $P(k,a) \approx P_{\rm lin}(k,a) \propto D_W^2(a)$ and so 
\be
 \frac{d\ln P(k,a)}{d\ln D_W(a)}\approx 2 \,.
 \label{eq:linear}
\ee
To determine the response of $D_W$ in \refEq{Rgrowth}, we solve \refEq{DW} with the initial
condition \refEq{D_ic}, and the result is shown in \reffig{dlnDW}. It approaches the matter
dominated ($\Omega_m=1$) limit $13/21$ at high redshift for both super-Jeans and sub-Jeans
scale responses, and at low redshift is smaller in the super-Jeans case as expected from
\reffig{dm}. In \reffig{dlnpk_linear}, we compare the measured power spectrum response in
sub-Jeans and super-Jeans SUs to the corresponding linear perturbation theory predictions,
i.e.~\refeqs{Rgrowth}{linear}. We find that in both cases the linear perturbation theory
agrees with the measured responses in the linear regime, i.e., at sufficiently low $k$ or
high $z$.

\begin{figure}[h]
\centering
\includegraphics[width=0.45\textwidth]{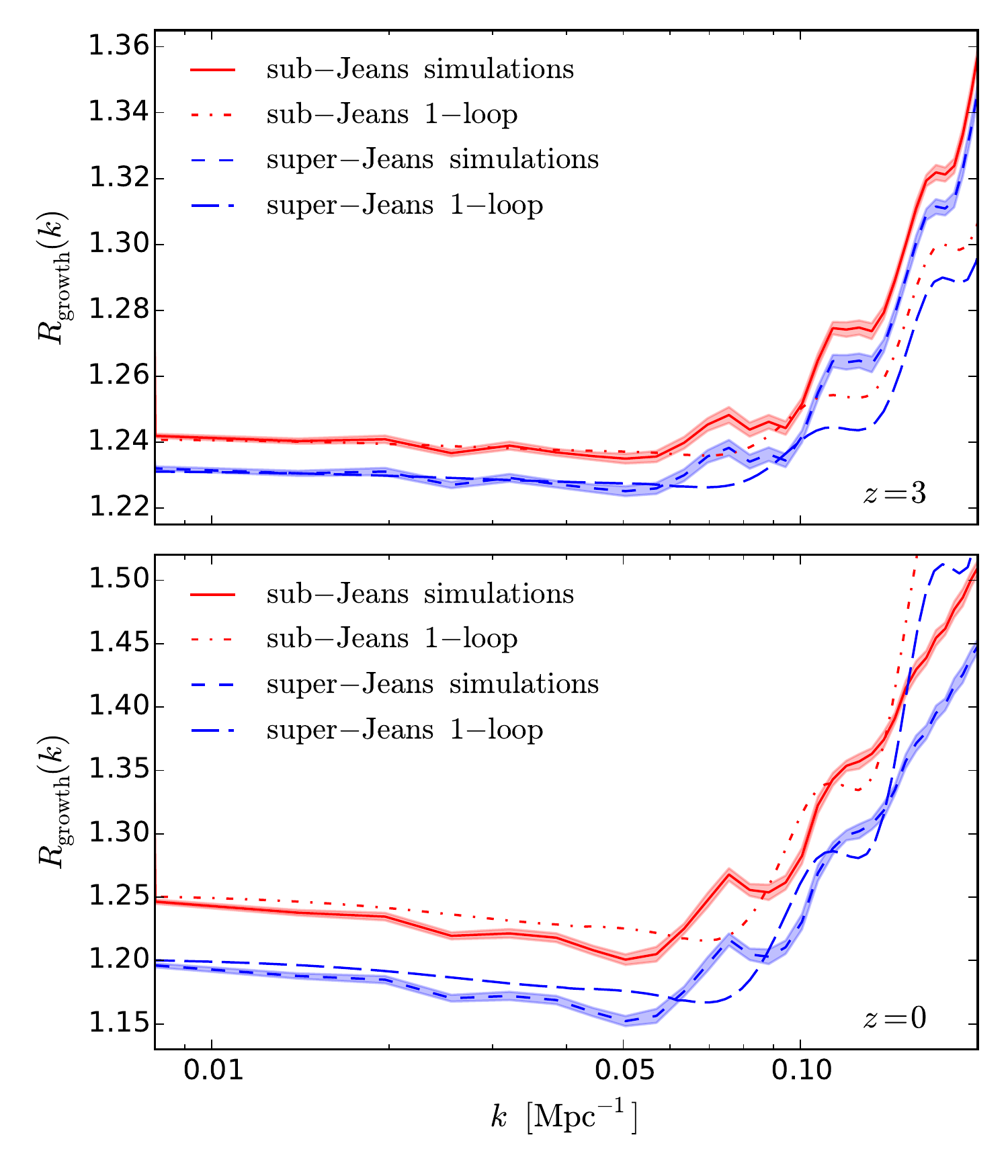}
\caption{Same as \reffig{dlnpk_linear}, but for the 1-loop predictions,
i.e. \refeq{Rgrowth} and \refeq{1loop}.}
\label{fig:dlnpk_1loop}
\end{figure}

However, as we move to lower redshift as well as higher $k$, the measured responses
become nonlinear and perturbation theory predictions deviate from SU simulation
measurements. Unlike perturbation theory, the SU response calibration is not limited
to large scales. On the other hand, we can understand the onset of nonlinearity in
the simulations through higher order perturbation theory. The 1-loop power spectrum
from standard perturbation theory is given by (see e.g. Ref.~\cite{Jeong:2006xd})
\be
 P_{\rm 1-loop}(k,a)=P_{\rm lin}(k,a)+P_{22}(k,a)+2P_{13}(k,a) \,,
\ee
where the nonlinear corrections $P_{22}$ and $P_{13}$ are proportional to $D_W^4$
if $\Omega_{mW}(a_W)/f_{W}^2(a_W)\approx1$.
Therefore,
\be
 \frac{d\ln P_{\rm 1-loop}(k,a)}{d\ln D_W(a)}
 =2\left[1+\frac{P_{22}(k,a)+2P_{13}(k,a)}{P_{\rm 1-loop}(k,a)}\right] \,,
 \label{eq:1loop}
\ee
which now is a function of $k$. Note that in the global cosmology $\Omega_m/f^2=1.034$
at $z=3$ and 1.203 at $z=0$, and the standard perturbation theory should work better
at $z=3$ than at $z=0$. Since the long-wavelength perturbation we consider is small
($|\delta_{\uparrow\downarrow0}|=0.01$), the standard perturbation theory should work
as well in both the sub-Jeans and super-Jeans SUs as the global universe.

In \reffig{dlnpk_1loop} we plot the 1-loop predictions in sub-Jeans (red dot-dashed)
and super-Jeans (blue long-dashed). We find that the 1-loop predictions extend the
agreement with the $N$-body measurement to smaller scales compared to the linear
predictions. More precisely, the difference between the 1-loop model and the measurement
at $z=3$ ($z=0$) is 1\% (3\%) at $k\sim0.1~{\rm Mpc}^{-1}$ and 4\% (6\%) at $k\sim0.2~{\rm Mpc}^{-1}$.
At even smaller scale or lower redshift, the nonlinearity is too large to be modeled
by the 1-loop perturbation theory. The SU simulation calibration technique itself is
not limited in wavenumber and our $N$-body implementation is instead only limited by
resolution as well as the lack of baryonic and astrophysical modeling in the deeply
nonlinear regime.

\subsection{Position-Dependent Power Spectrum}
\label{sec:ibn}
The power spectrum response can also be tested in simulations and observed in surveys 
through the position-dependent power spectrum. As a simulation based test, it also serves
to check the SU calibration of the power spectrum response deep into the nonlinear regime.

Specifically we compare the response measured from the small-box SU simulations to the
squeezed-limit position-dependent power spectrum measured from the big-box simulations
with the global cosmology (without a uniform long-wavelength density fluctuation). In
the latter, we assume that the dark energy does not cluster with matter and so its
position-dependent power spectrum should match the sub-Jeans SU prediction.

The procedure of measuring the position-dependent power spectrum is explained in detail
in \cite{Chiang:2014oga}. In short, we first distribute the dark matter particles onto
a $2048^3$ grid by the CIC density assignment scheme to construct the density fluctuation.
We next divide the big-box simulations in each dimension by 8, so there are  $N_s=512$
subvolumes in total  with  comoving side length of $L=350$ Mpc. In each subvolume centered at
${\mathbf r}_L$, we measure the local power spectrum as $\hat{P}(k,a|{\mathbf r}_L)$
and the mean overdensity (with respect to the entire box) as $\hat{\bar{\delta}}_m({\mathbf r}_L)$
and construct 
\be
 \frac{\frac{1}{N_s}\sum_{{\mathbf r}_L}\hat{P}(k,a|{\mathbf r}_L)\hat{\bar{\delta}}_m({\mathbf r}_L)}
 {\left[\frac{1}{N_s}\sum_{{\mathbf r}_L}\hat{P}(k,a|{\mathbf r}_L)\right]
 \left[\frac{1}{N_s}\sum_{{\mathbf r}_L}\hat{\bar{\delta}}_m^2({\mathbf r}_L)\right]} \,,
\ee
where the summation is over the 512 subvolumes in one big-box realization. The correlation
between $\hat{P}(k,a|{\mathbf r}_L)$ and $\hat{\bar{\delta}}_m({\mathbf r}_L)$ quantifies
the integrated bispectrum, and in the squeezed limit where $k\ll1/L$ the integrated bispectrum
can be understood as the {\it total} response of the power spectrum to the long-wavelength
overdensity.

\begin{figure}[h]
\centering
\includegraphics[width=0.48\textwidth]{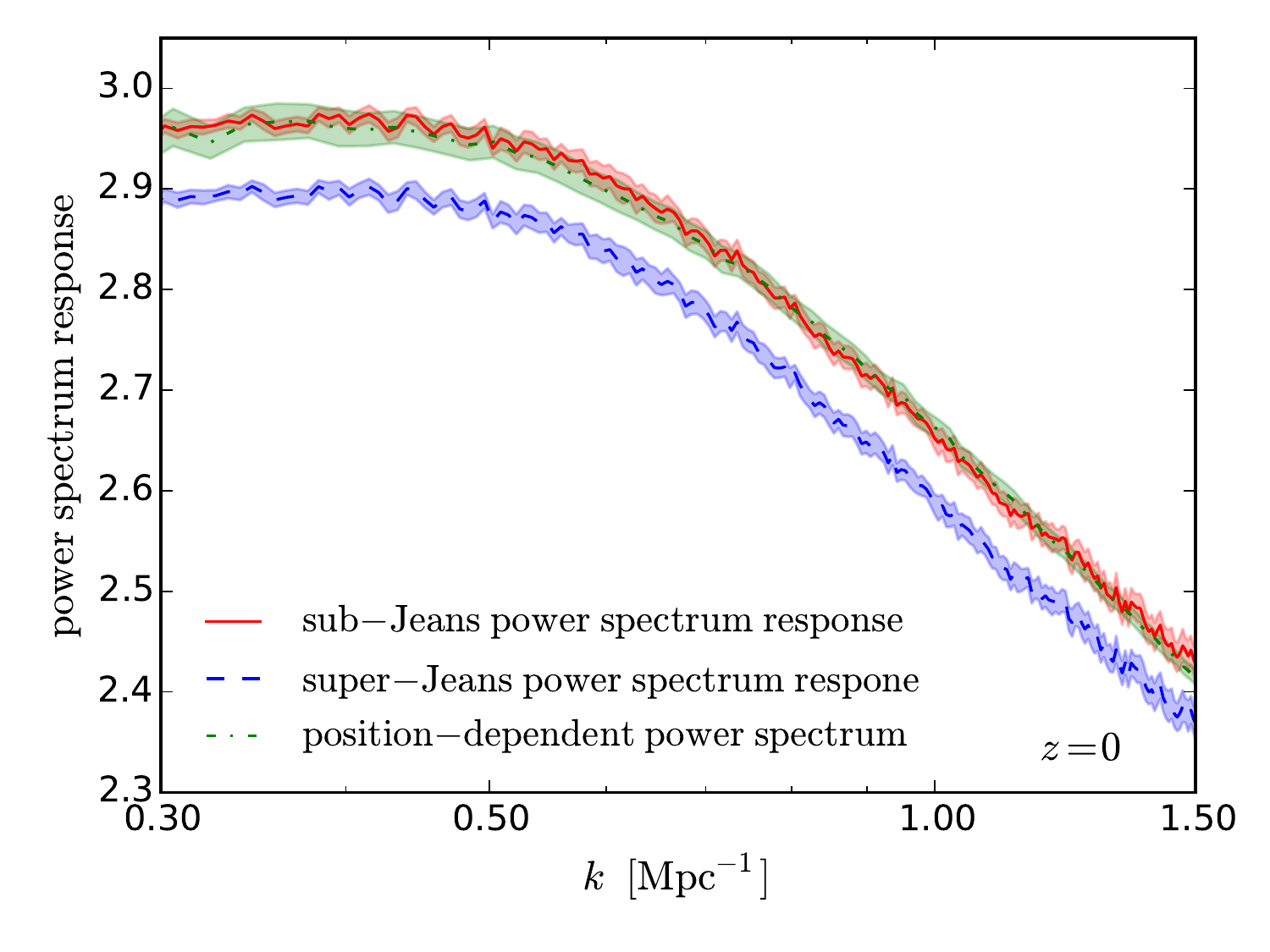}
\caption{Comparison of the squeezed-limit position-dependent power spectrum of big-box,
global simulations (green dot-dashed), and the total power spectrum response of sub-Jeans
(red solid) as well as super-Jeans (blue dashed) separate universe simulations at $z=0$.
The shaded areas show the error on the mean. This figure summarizes our main power spectrum
response results: the position-dependent power spectrum and the sub-Jeans power spectrum
response agree significantly better than the difference in response across the Jeans scale
confirming the scale dependence in this observable deep into the nonlinear regime.}
\label{fig:ibn}
\end{figure}

\refFig{ibn} shows the comparison at $z=0$ between the total power spectrum response from
the previous section (red solid for sub-Jeans and blue dashed for super-Jeans cases) and
the position-dependent power spectrum  (green dot-dashed) averaged over the 20 realizations
with its error. To reach the squeezed limit, we require $k\gtrsim100/L\sim0.3~{\rm Mpc}^{-1}$,
and in this regime the agreement with the SU response is better than a percent. This agreement
is significantly better than the difference between the super-Jeans and sub-Jeans power spectrum
responses and thus verifies the SU calibration technique. With the SU technique tested into
the nonlinear regime, we can apply these results to the super-Jeans case of the position-dependent
power spectrum without the need for costly simulations that include dark energy clustering.

\section{Scale-Dependent Halo Bias}
\label{sec:bias}
The SU simulations also calibrate the response of the halo mass function to a long-wavelength 
mode and hence the bias of the halo number density due to that mode. In \refsec{responsebias},
we review the technique for measuring halo bias from SU simulations and show that in the
quintessence model it acquires a scale dependence at the Jeans scale. In \refsec{clusteringbias},
we test this response bias in the SU simulations against the clustering bias extracted from
40 small-box global simulations. We discuss the implications of scale dependence for the
temporal nonlocality of halo bias and the observability of features in the halo power
spectrum in \refsec{interpretation}.

\subsection{Response Bias}
\label{sec:responsebias}

The linear density bias $b_1(M)$ of halos of mass $M$ can be defined as the response
of the differential halo abundance $n_{\ln M}=dn/d\ln M$ to the long-wavelength mode
\be
 b_1(M) \equiv \frac{d\delta_h}{d\delta_m} = \frac{d\ln n_\lnM}{d\delta_m} \,,
 \label{eq:biasasresponse}
\ee
which we call ``response bias''. Thus by measuring the response of the halo mass
function in the SU simulations we have a direct calibration of response bias
\cite{Li:2015jsz,Lazeyras:2015lgp,Baldauf:2015vio}. Note that the derivative in
\refeq{biasasresponse} is evaluated at a fixed time, but will depend on the whole
growth history of $\delta_m(a)$. This temporal nonlocality implies that response
bias can be scale dependent if that growth history is also scale dependent. For
quintessence, the SU simulations allow us to calibrate the bias above and below
the Jeans length of quintessence without direct simulations of its clustering properties.

As discussed in Ref.~\cite{Li:2015jsz}, response bias largely reflects the change
in the masses of halos due to the same local change in growth that affects the power
spectrum. The enhanced growth in $\delta_m>0$ regions makes halos more massive locally
than in their $\delta_m<0$ counterparts. Hence halos of a fixed mass are associated
with the more abundant lower peaks in the initial density field in the former and
the less abundant higher peaks in the latter. This also means that measuring the
change in abundance of halos in fixed mass bins between the SU simulation pairs is
an inefficient way to quantify response bias. Halos with small changes in mass across
wide mass bins would register their response only when individual halos move across
mass bins.

We instead adopt abundance matching as introduced in Ref.~\cite{Li:2015jsz}, which
we now summarize. By finding the mass threshold above which the cumulative abundance
is fixed, we largely eliminate the sampling noise from the discrete nature of halos.

\begin{figure}[h]
\centering
\includegraphics[width=0.45\textwidth]{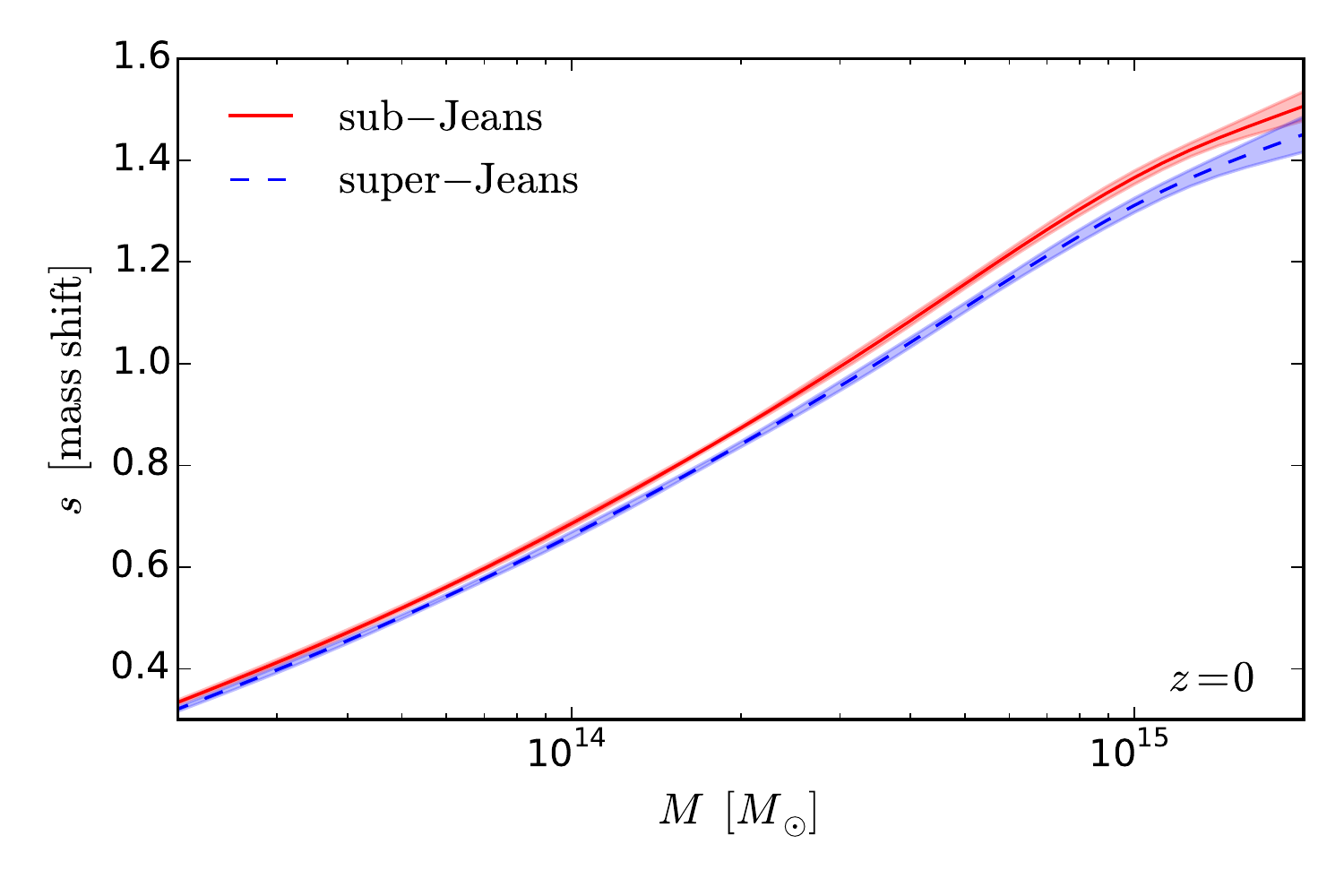}
\caption{Threshold mass shift as a response of varying $\delta_m$ at fixed cumulative abundance
at $z=0$. The solid line and shaded region show the smoothed estimate and the bootstrap error.}
\label{fig:s_z0}
\end{figure}

Specifically for either the sub-Jeans or the super-Jeans case, we first combine halo catalogs
of all realizations of the same $\delta_m$ in the small-box suite. The masses of the $i^{\rm th}$
most massive halo $M_i^\pm$ from the $\delta_m = \pm 0.01$ SU simulations determine the
discrete threshold mass shift
\be
 s_i(\lnM_i) = \frac{\lnM_i^+ - \lnM_i^-}{2|\delta_m|} \,,
\ee
where $M_i$ is the geometric mean of $M_i^+$ and $M_i^-$. We then use the smoothing spline
technique to estimate the ensemble average threshold mass shift $\hat s(\ln M)$ as well as
the cumulative halo abundance above threshold mass $\hat n(\ln M)$. \refFig{s_z0} shows the
mass shift measured  from 20 sub-Jeans and super-Jeans SU simulations at $z=0$ as a function
of halo mass. We find that the mass shift due to varying $\delta_m$ is smaller above the
Jeans scale, which reflects the fact that its growth history makes it closer to the global
universe than below the Jeans scale.

\begin{figure}[h]
\centering
\includegraphics[width=0.46\textwidth]{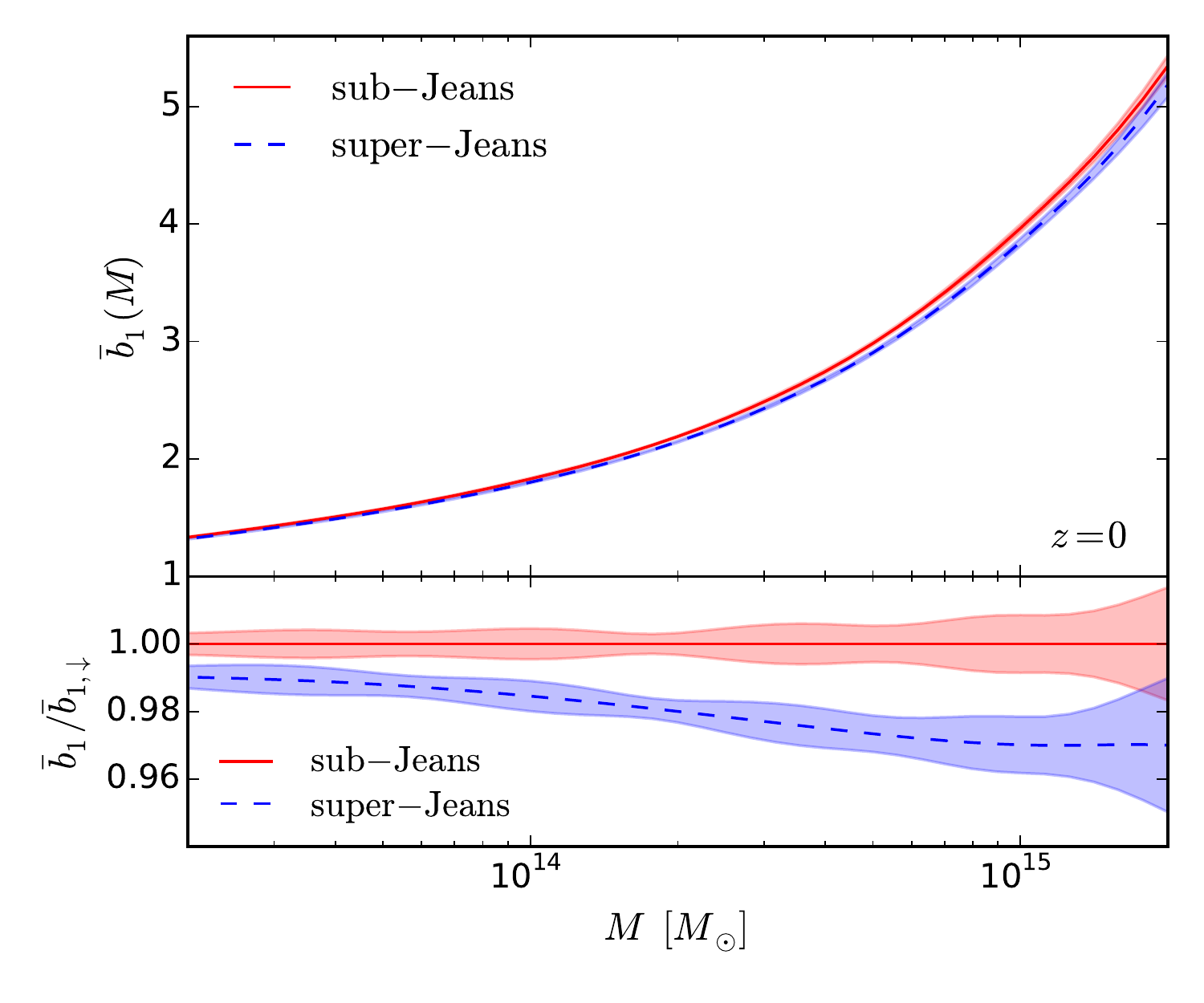}
\caption{(Top) The $z=0$ response biases measured from 20 sub-Jeans (red solid) and
super-Jeans (blue dashed) separate universe simulations. The lines and shaded areas
show the smoothed estimate and the bootstrap error. (Bottom) The ratios of the response
biases to that of the sub-Jeans response bias. The difference between the sub-Jeans
and super-Jeans response biases, which indicates that the linear halo bias is scale
dependent in the presence of the scale-dependent growth, is the second central result
of our separate universe simulations.}
\label{fig:b_resp}
\end{figure}

The halo mass function follows as the derivative of the cumulative mass function
$\hat n_\lnM=-d\hat n/d\lnM$. We can then estimate the Lagrangian halo bias above
threshold mass $M$ as
\be
 \hat{\bar{b}}_1^\Lr(M) = \frac{\hat n_\lnM(\ln M)\,\hat s(\ln M)}{\hat n(\ln M)} \,.
\label{eq:am}
\ee
This quantity is the Lagrangian bias since the SU simulations are performed with the
same comoving rather than physical volume. The dilation of the volume from the change
in scale factors brings the cumulative Eulerian bias to
\be
 \hat{\bar{b}}_1(M) =1+ \hat{\bar{b}}_1^\Lr(M)\,.
 \label{eq:eulerianb}
\ee
In \reffig{b_resp} we compare the response bias on super-Jeans (blue dashed) and sub-Jeans 
(red solid) scales as a function of halo mass at $z=0$. The bias at a fixed mass is smaller
in the super-Jeans case. Just like for the power spectrum response, above the Jeans scale
for the same final $\delta_m$, the SU is closer to global at high redshift. Thus the change
in growth and the consequent change in halo masses and abundances is smaller. We find that
the mild mass dependence of the fractional difference between the super-Jeans and sub-Jeans
response biases is due mainly to the dilation effect in \refeq{eulerianb}, since that of
the Lagrangian bias is fairly mass independent due to the similar shapes of the mass shift
displayed in \reffig{s_z0} (see also \reffig{bLversusmodels}).

\subsection{Clustering Bias}
\label{sec:clusteringbias}

To verify the SU calibration of halo bias through the mass function response, we can
compare it to how linear halo bias is commonly measured from the two-point statistics,
which we call clustering bias
\be
\bar b_1(M) = \lim_{k\to0}\frac{P_{hm}(k;M)}{P_{mm}(k)} \,,
 \label{eq:biasascrosspower}
\ee
where $P_{hm}$ is the cumulative halo number density cross power spectrum with the
matter density. Where no confusion should arise, we omit the $M$ argument of the
cumulative bias. Above the Jeans scale of quintessence, this approach would require
simulations of quintessence clustering even for linear halo bias. Below the Jeans
scale, we can test the equivalence of response and clustering bias with global
simulations where quintessence enters only at the background level.

In order to extract the $k\rightarrow 0$ limit, we first compute
\be
 \bar{q}(k)=\frac{{P}_{hm}(k)}{P_{mm}(k)} \,,
\ee
for each of the 40 simulations of the global cosmology for a set of mass thresholds.    
Motivated by Ref.~\cite{Assassi:2014fva}, we fit $\bar{q}(k)$ to the model
\be
 \bar{b}(k)=\bar{b}_1+\sum_{i=1}^n \bar{b}_{k^{2n}}k^{2n} \,,
\ee
where we treat $\bar{b}_{k^{2n}}$ as nuisance parameters that absorb the loop
corrections in the large-scale limit. We then get the best-fit bias parameters
by minimizing
\be
 \chi^2=\sum_{k}^{\kmax}\frac{[\bar{q}(k)-\bar{b}(k)]^2}{\sigma^2[\bar{q}(k)]} \,,
\label{eq:chi2}
\ee
where $\sigma^2[\bar{q}(k)]$ is the variance of $\bar{q}(k)$ measured from 40 global
small-box simulations.

To ensure the robustness of the fitted clustering bias, especially as compared with the
small predicted difference between sub-Jeans and super-Jeans response biases, we examine
the bias models with $n=0$, 1, and 2 for various $\kmax$. We seek consistent result for
different bias models (different $n$) and $\kmax$. The general principle is that the
larger the $\kmax$, the larger the $n$ required to account for the nonlinearity and to
avoid underfitting. Conversely, for models with $n>0$ $\kmax$ cannot be too small or the
fit would suffer from overfitting.

With each bias model and  $\kmax$, we visually inspect its goodness of fit to $\bar{q}(k)$
for various threshold halo masses. We find that across two decades in halo mass
($2\times10^{13}-2\times10^{15}\,M_\odot$), the bias models of $n=0$, 1, and 2 with the
biases fitted to $\kmax=0.014-0.028\,{\rm Mpc}^{-1}$, $0.042-0.049\,{\rm Mpc}^{-1}$,
and $0.056-0.07\,{\rm Mpc}^{-1}$ are in agreement with the mean $\bar{q}(k)$, and the
agreement even extends to $k>\kmax$. This shows that the fit is free from overfitting
and underfitting problems. For a given halo mass, the best-fit clustering bias varies
up to 0.2\%, 0.5\%, and 2\% among different $n$ and $\kmax$ at $2\times10^{13}\,M_\odot$,
$2\times10^{14}\,M_\odot$, and $2\times10^{15}\,M_\odot$, respectively. Given the fact
that the clustering bias is stable for various bias models and fitting range, we conclude
that systematic error due to $n$ and $\kmax$ is at most comparable to our statistical
error, and is the largest at the high-mass end at which the statistical error is also large.

\begin{figure}[h]
\centering
\includegraphics[width=0.48\textwidth]{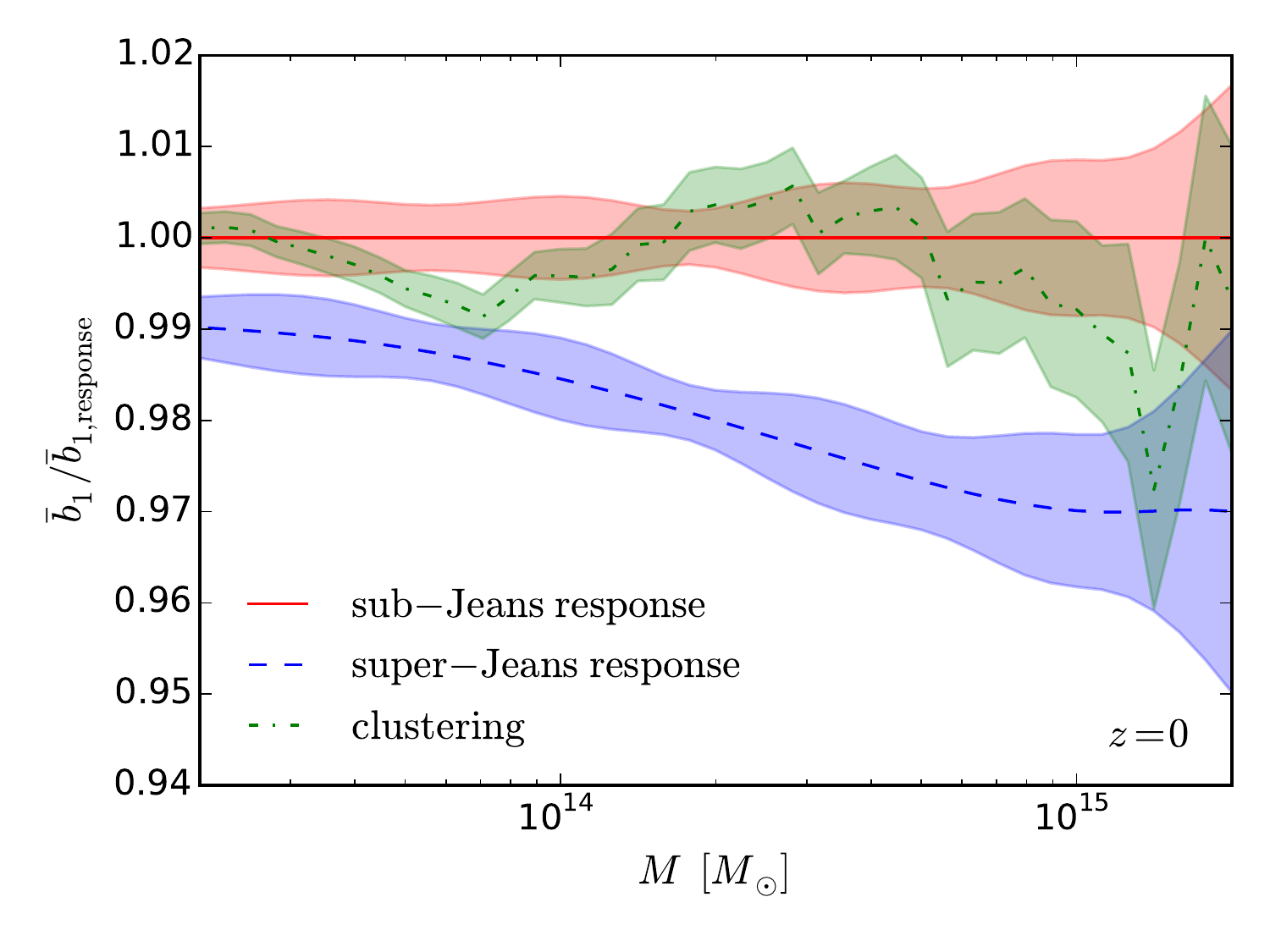}
\caption{The ratios of the linear biases to that of the sub-Jeans response bias at $z=0$.
The red solid and blue dashed lines show the sub-Jeans and super-Jeans response biases
measured from 20 separate universe simulations, whereas the green dot-dashed line shows
the clustering bias measured from 40 global simulations. The error of the clustering bias
is measured from the scatter of the 40 simulations. This figure summarizes the main results
on halo bias: the agreement between the clustering bias measured from global simulations
and the sub-Jeans response bias verifies the observable difference in halo bias across the
Jeans scale inferred from the separate universe simulations.}
\label{fig:b_clus}
\end{figure}

In \reffig{b_clus} we shows our fiducial results of the clustering bias measurement for
the quadratic model ($n=1$) with $\kmax=0.49\,{\rm Mpc}^{-1}$, which gives the smallest
statistical errors. We find that the clustering bias is in good agreement with the sub-Jeans
response bias across two decades in halo mass, confirming the validity of the SU technique.
This agreement is substantially better than the difference between the super-Jeans and sub-Jeans
response bias at low- and mid-mass regime, even after including the systematic differences
between the fitting techniques.

To further test robustness of the scale-dependent bias result, we also try the halo
finding algorithm provided in Ref.~\cite{Li:2015jsz}, another spherical overdensity finder
similar to that in Ref.~\cite{Tinker:2008ff}. We find that the clustering bias is statistically
in equally good agreement with the sub-Jeans response as well.

With this verification of the SU calibration of halo bias, our results represent the first
simulation confirmation of scale-dependent halo bias from scale-dependent growth. A related
effect on the void bias has been measured in the simulations with cold dark matter and massive
neutrinos \cite{Banerjee:2016zaa}.

\begin{figure}[h]
\centering
\includegraphics[width=0.45\textwidth]{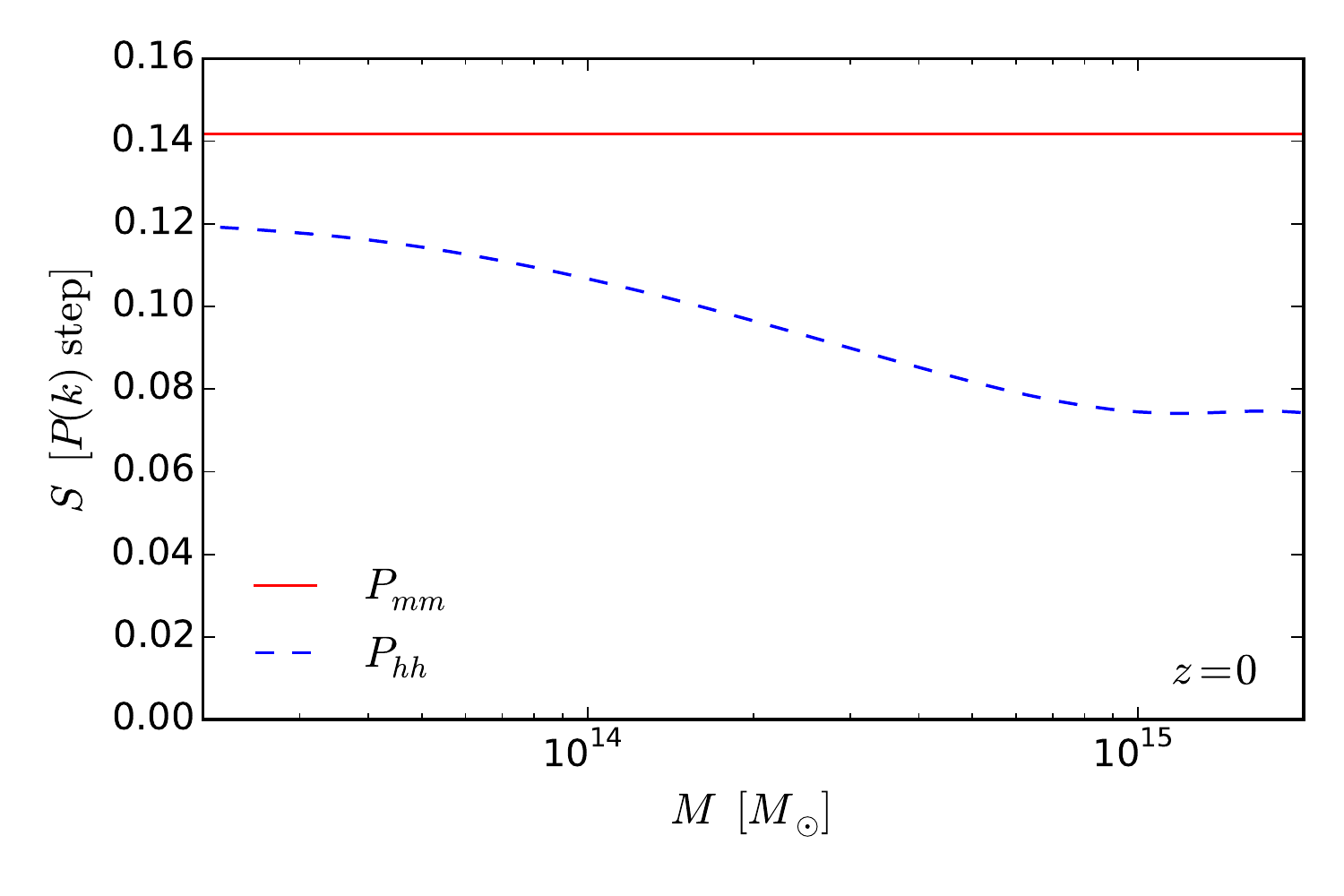}
\caption{Fractional difference or ``step'' across the Jeans scale in the matter (red solid,
\refeq{Smm}) versus the halo (blue dashed, \refeq{Shh}) power spectra. The step in the halo
power spectrum is reduced by approximately a factor of 2 compared with the matter power
spectrum at high mass where the Lagrangian bias contribution dominates. Bias prescriptions
which assume locality at either the observed or initial redshift predict the same step or
zero step respectively at high masses.}
\label{fig:phh_ratio}
\end{figure}

\subsection{Scale-Dependent Bias and Power Spectra}
\label{sec:interpretation}

Since the halo bias is smaller above versus below the Jeans scale of quintessence, its
scale dependence counters the growth rate effects in the matter power spectrum. This
is especially true at high masses where the Lagrangian bias dominates. The change in
the linear growth function above $(D^\uparrow)$ versus below $(D^\downarrow)$ the Jeans
scale leads to a step in the linear matter power spectrum of approximately
\begin{equation}
\label{eq:Smm}
S_{mm} \equiv 
2 \frac{D^\uparrow - D^\downarrow}{D^\downarrow}\,,
\end{equation}
whereas the step in the cumulative halo power spectrum is
\begin{equation}
S_{hh}
 \equiv
  2 \frac{D^\uparrow \bar b_1^\uparrow - D^\downarrow \bar b_1^\downarrow}{D^\downarrow \bar b_1^\downarrow} \,.
  \label{eq:Shh}
\end{equation}
In \reffig{phh_ratio}, we show the amplitude of these steps as a function of mass.
At the high mass end the halo power spectrum has half the step amplitude of the
matter power spectrum.

This result not only confirms that scale-dependent linear growth leads to scale-dependent
bias, but it does so in a way that both reduces the observability of features in the halo
power spectrum \cite{LoVerde:2014pxa} and violates principles that underlie simple models
for bias. It is commonly assumed that the statistics of halos at any observation epoch is
determined solely by the statistics of the linear density field at a single epoch, and
hence bias is scale-free with respect to the matter power spectrum at that epoch. For
models with scale-dependent growth, this epoch is commonly taken to be  the initial epoch
for models where the growth becomes scale-free during matter domination \cite{Parfrey:2010uy}.

For example in the excursion set, Lagrangian bias is given by the conditional probability
that the initial density field crosses some barrier at a smoothing scale $R_S$ corresponding
to the mass $M$ at the background density given that it takes the value $\delta_m$ at some
larger scale $R_L$ via a random walk between the two. For our quintessence case where these
scales are arbitrarily well separated by our SU assumption, the lack of correlation in the
Gaussian random initial conditions between these scales means that halo bias is local in the
initial density field. Specifically, the conditional probability cannot depend on steps in
the random walk with $R>R_L$, and hence whether $\delta_m$ was achieved from super-Jeans or
sub-Jeans scale fluctuations. This holds regardless of the shape of the barrier, its dependence
on the redshift of observation, or some putative intermediate epoch of halo formation. As
a result, Lagrangian halo bias should be scale-free with respect to the matter power spectrum
at the initial epoch.

As emphasized in Ref.~\cite{Parfrey:2010uy}, even if the Lagrangian bias is scale-free with
respect to the initial power spectrum, it becomes scale dependent with respect to the matter
power spectrum at the observation epoch. This effect is solely due to the scale-dependent
growth in the latter, and hence the scale dependence takes a simple and specific form
$b_1^{L\uparrow} = (D^\downarrow/D^\uparrow) b_1^{L\downarrow}$. In \refapp{bias_model},
we review this construction in more detail. In the quintessence model this means that the
step in the halo power spectrum should be absent when the Lagrangian bias dominates
$\lim_{M\rightarrow \infty}S_{hh}=0$ (see \refeq{Shh}). Our results significantly violate
this prediction.

Similarly models of halo bias that rely on a universal mass function ansatz, characterize
the bias as its derivative with respect to $\delta_m$ at the observation epoch. If this
derivative depends {\it only} on the local density field $\delta_m$ at the observation epoch,
for example by assuming a change in the spherical collapse threshold $d\delta_c/d\delta_m=-1$
(see \refapp{bias_model}), then the Lagrangian bias would be local and hence scale-free with
respect to the matter power spectrum at the observation epoch. Our results for scale-dependent
bias directly violate this prediction and are essentially half-way between these two extreme
models. We find that halo bias is nonlocal in time and cannot be characterized by the statistics
of the density field at a single epoch, initial or observed when there is scale-dependent
linear growth.

In \refapp{bias_model}, we show that encapsulating the dependence on the growth history of
$\delta_m(a)$ through its impact on the spherical collapse threshold at the observation epoch
and assuming a universal mass function characterizes the quintessence SU simulation results
better than either of these simplistic models. However given the assumptions underlying this
type of modeling, its validity in other contexts should be tested directly in simulations.

\section{Discussion}
\label{sec:discussion}

Quintessence dark energy provides an arena to explore the response of small-scale
observables to the amplitude, scale, and growth history of long-wavelength fluctuations
with the separate universe technique. In the presence of quintessence fluctuations,
the growth of long-wavelength fluctuations differ above and below its Jeans scale.   
We verify that even below the Jeans scale, where a naive separate universe picture
does not strictly apply because the local curvature evolves due to non-gravitational
forces which keep the quintessence smooth, the response of small-scale observables
can still be accurately modeled by a modified expansion history alone. One implication
of this finding is that halo bias is not directly a response of halo number density
to the local curvature, but rather to the local expansion history.

Using this technique, we show that in the presence of the scale-dependent growth,
the local power spectrum and halo mass function acquire a dependence on the scale
of the long-wavelength mode. Equivalently, the squeezed bispectrum and halo bias
become scale dependent. To our knowledge, our results are the first verification
of scale-dependent bias from scale-dependent growth using simulations. Moreover
they violate predictions of models where bias is effectively local in the density
field at a single epoch, initial or observed, and show that halo bias is temporally
nonlocal. Likewise the nonlinear matter power spectrum cannot simply be a function
of the linear power spectrum at the same epoch.

Specifically, we use the separate universe (SU) technique to perform $N$-body
simulations in the sub-Jeans and super-Jeans SUs. By differencing pairs of
overdense and underdense SU simulations with the same Gaussian realizations
of initial phases, much of the sample variance is canceled, and so we can
precisely characterize the responses of the power spectrum (which is equivalent
to the squeezed-limit bispectrum) and the halo mass function (which gives the
linear halo bias) to the long-wavelength matter fluctuation.

We validate the SU approach by comparing to perturbation theory predictions
for the power spectrum response in both the super-Jeans and sub-Jeans limits
(see \reffigs{dlnpk_linear}{dlnpk_1loop}). Since it is the sub-Jeans limit
where the SU technique might naively fail, we further test it with direct
simulations that possess long-wavelength matter modes in big-box simulations
with smooth dark energy. We find that the squeezed-limit position-dependent
power spectrum measured from the big-box simulations agrees with the power
spectrum response to the resolution limit $k\sim1\,{\rm Mpc}^{-1}$. Similarly,
the clustering bias is statistically consistent with the response bias across
two decades in halo mass ($\sim\!10^{13}\!-\!10^{15}\,M_\odot$). Thus, with
the SU technique verified into the nonlinear regime, we can robustly assess
the scale-dependence of the power spectrum and halo density responses across
the Jeans scale without costly simulations that include quintessence clustering.

We show that for both responses there is a statistically significant distinction
between sub-Jeans and super-Jeans SUs at $z=0$. More precisely, the power spectrum
response in the super-Jeans SU is roughly 2\% smaller than that in the sub-Jeans
SU for $k\lesssim1\,{\rm Mpc}^{-1}$; the halo bias in the super-Jeans SU is roughly
1\% and 3\% smaller than that in the sub-Jeans SU for halo mass of $2 \times 10^{13}$
and $2\times 10^{15}~M_\odot$ respectively. The fact that the response is smaller
in the super-Jeans SU is because quintessence enhances the growth of matter
fluctuations there, and so the super-Jeans overdensity was smaller in the past.
These key SU results, along with the comparison to the global simulations,
are summarized in \reffig{ibn} and \reffig{b_clus}.

More generally, this dependence on the growth history of the long wavelength fluctuation
indicates that the response of small scale observables is nonlocal in time. In particular,
the statistically significant difference between sub-Jeans and super-Jeans response biases
measured in our SU simulations falsifies the standard Lagrangian picture where the statistics
of halos at any observation epoch is determined solely by the statistics of the linear density
field at a single epoch.

These effects are in principle important for interpreting observational tests of quintessence
clustering from galaxy surveys and their cross correlation with the CMB (e.g. \cite{Bean:2003fb,Hu:2004yd}).
In particular, the step feature in the halo power spectrum is smaller by up to a factor of 2
compared with the matter. However these corrections, while significant relative to the clustering
effects on the matter power spectrum itself, are small in an absolute sense for observationally
viable dark energy equations of state (i.e. $w_Q\approx-1$) in the absence of quintessence
isocurvature fluctuations \cite{Gordon:2004ez,Hu:2016ssz}.

On the other hand, the same technique which has been validated here using quintessence,
can be applied to more observationally viable cosmological models, such as those with
massive neutrinos. Massive neutrinos cluster with dark matter on large scales, but their
free streaming sets an effective Jeans scale. This would generate not only a feature in
the two-point function of the total matter \cite{Lesgourgues:2006nd}, but also influence
the high-order statistics \cite{Shoji:2009gg,Blas:2014hya,Fuhrer:2014zka,Levi:2016tlf}
as well as the halo bias \cite{LoVerde:2014pxa}. We intend to apply the SU technique
to study how massive neutrinos affect the small-scale structure formation in a
future work.

\acknowledgements{
We thank Eiichiro Komatsu and Fabian Schmidt for useful discussions.
We would also like to thank Alexander Knebe for guiding us to implement the dark energy model into Amiga Halo Finder. 
WH thanks the Aspen Center for Physics, which is supported by National Science Foundation grant PHY-1066293, 
where part of this work was completed.
Results in this paper were obtained using the high-performance computing system
at the Institute for Advanced Computational Science at Stony Brook University
and with the computation and storage resources
provided by the University of Chicago Research Computing Center.
CC and ML are supported by grant NSF PHY-1316617.
WH was supported by U.S.~Dept.\ of Energy contract DE-FG02-13ER41958,
NASA ATP NNX15AK22G, and  the Kavli Institute for Cosmological Physics
at the University of Chicago through grants NSF PHY-0114422
and NSF PHY-0551142.}

\appendix

\section{Bias model comparisons}
\label{app:bias_model}
Here we make brief comparisons with formalisms for scale-dependent halo bias
in the recent literature. Analytic models for scale-dependent bias arising
from scale-dependent growth exist and have been applied to modified gravity
\cite{Hui:2007zh, Parfrey:2010uy} and to massive neutrino cosmologies \cite{LoVerde:2014pxa, LoVerde:2016ahu}.
Both formalisms can be applied to the quintessence model studied in this paper.  

\subsection{PHS11: Excursion set on the initial density field}
Parfrey, Hui, and Sheth \cite{Parfrey:2010uy} (hereafter PHS11) develop the theory
of excursion sets for a cosmology with a scale-dependent linear growth function
$D(z;k)$ caused by modified gravity associated with the late-time accelerated
expansion of the universe. We can equally well apply their construction to the
quintessence model.

In the PHS11 model, halos are identified by performing a barrier-crossing calculation
on the initial density field, $\delta_i$ at some redshift in the matter dominated era,
$z_i$, {\em before} the linear growth of the matter field becomes scale dependent.
In the PHS11 model the scale-dependent evolution of density perturbations (i) causes
the barrier to be scale-dependent (ii) induces scale-dependent Lagrangian bias from
the scale-dependent mapping between the statistics of the density field at $z_i$ and
those at a later redshift $z$. 
 
As discussed in \refsec{interpretation}, for the quintessence model a scale-dependent
barrier can change the mass dependence of the bias but cannot introduce a scale dependence.
Unlike the modified gravity model of PHS11, we assume in the quintessence model an
arbitrarily large separation of scales involved in the barrier crossing and the
long-wavelength modes enforced by the SU approximation. In addition PHS11 find that
in their model the approximation of a  flat barrier (constant with scale) is good.
Their model for the Lagrangian bias observed at redshift $z$ is
\be
 b_1^L(z; k) = \frac{D(z_i;k)}{D(z;k)}b_1^L(z_i) \,,
\label{eq:PHSbL}
\ee
where the Lagrangian bias at $z_i$ is independent of $k$ and for a flat barrier $\delta_c$
is explicitly
\be
 b_1^L(z_i)= \left[ \frac{\delta_{c}^2}{\sigma^2(z_i)}-1\right]\frac{1}{\delta_{c}} \,.
\ee
The barrier height is typically derived from the spherical collapse model with
$\delta_{c}(z_{\rm form})$ as the critical value of an initial overdensity that will
collapse to ``form'' a halo by $z_{\rm form}$. The bias in PHS11 is a function of both
the observed $z$ and $z_{\rm form}$ and the bias at $z=z_{\rm form}$ can be scale dependent.
While these distinctions can change the prediction for the value of the bias as a
function of  mass, they do not introduce an additional dependence on $k$.

\begin{figure}[t]
\centering
\includegraphics[width=0.45\textwidth]{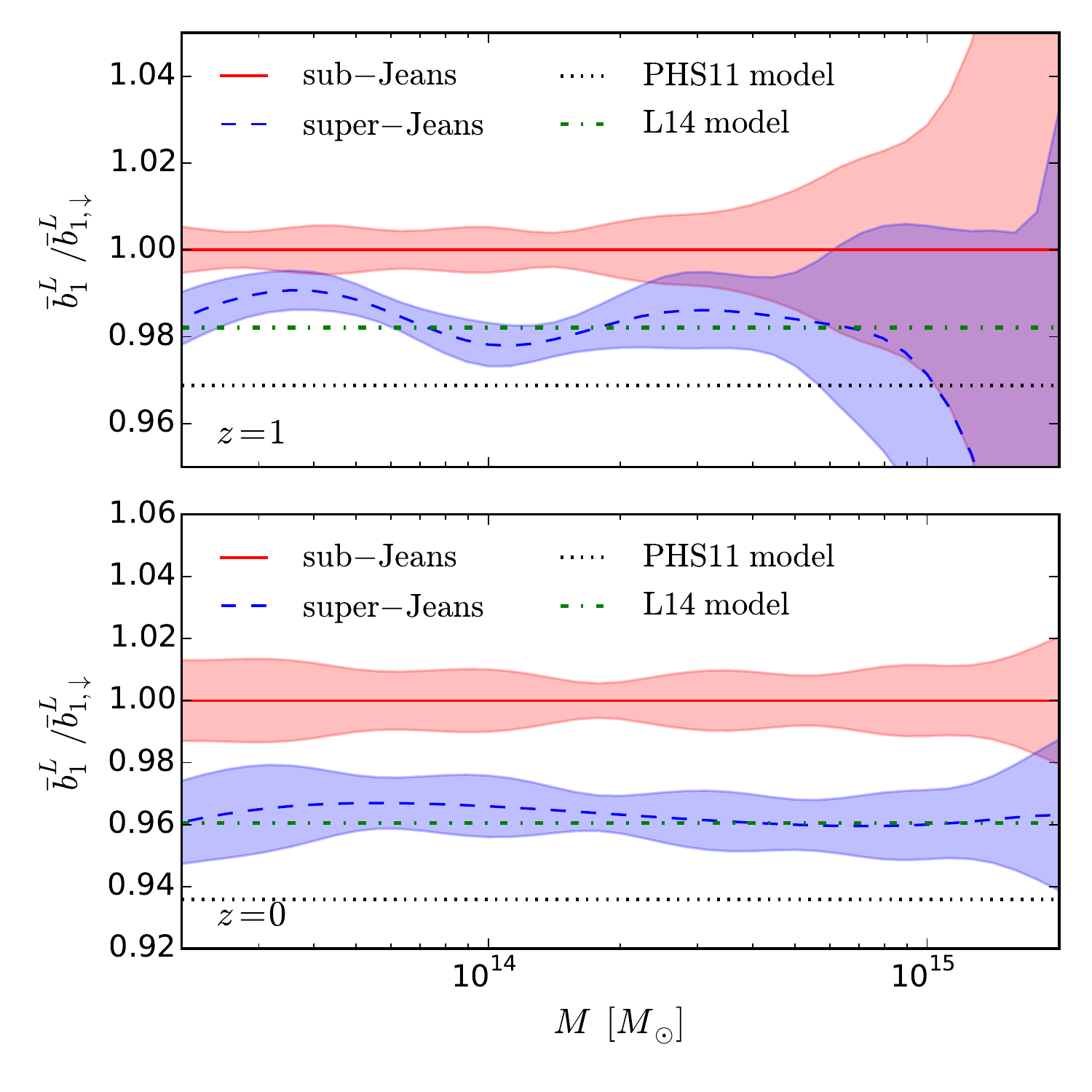}
\caption{Ratio of the super-Jeans to sub-Jeans scale Lagrangian bias factors at $z=1$
(top) and $z=0$ (bottom) measured from separate universe simulations in comparison with
the two models, PHS11 in \refeq{PHSbL} and L14 in \refeq{L14}, discussed in this Appendix.
Oscillations are mainly due to spline smoothing. Increasing the number of knots gives
flatter curves with larger errors but consistent averages. Large errors at $z=1$ for
$M\gtrsim 10^{15} M_\odot$ are due to the lack of high redshift massive halos. The PHS11
model predictions rely on the standard assumption that Lagrangian bias is scale-independent
with respect to the initial density field, and our simulations rule this out.}
\label{fig:bLversusmodels}
\end{figure}

The PHS11 flat-barrier expression for the bias in \refeq{PHSbL} has a particularly
clean interpretation: halos have scale-independent Lagrangian bias with respect to
the initial matter power spectrum
\ba
 P^L_{hh}(k;z) \:&\equiv \left[b_1^L(z; k)\right]^2 P_{mm}(k;z) \\
 \:&= [b_1^L(z_i)]^{2}P_{mm}(k;z_i)
 \,. \nonumber
\ea
The scale dependence of the Lagrangian halo bias arises entirely from the evolution of
the matter power spectrum leaving no scale in the Lagrangian halo power spectrum. We
test this concrete and robust prediction in the main text and find that the quintessence
model significantly violates this expectation. This point is further illustrated in
\reffig{bLversusmodels}, which compares the fractional change in the Lagrangian bias
measured in our simulations to the prediction in \refeq{PHSbL}. The flaw in this approach, 
that bias is local in the initial density field, can be ameliorated by making the barrier
explicitly dependent on the random walk on larger scales. This is discussed in the next
section.

\subsection{L14: Peak background split on a universal mass function}

A related but conceptually distinct calculation for scale-dependent bias, developed for
massive neutrinos, is given in LoVerde \cite{LoVerde:2014pxa} (hereafter ``L14").
The L14 model makes the assumption that the halo abundance at $z$ is given by a universal
function of $\nu \equiv\delta_c(z;k)/\sigma(M, z)$. Here $\delta_{c}(z;k)$ is associated
with the spherical collapse threshold at $z$ computed in the presence of the long-wavelength
or ``background'' mode $\delta_{m}(k)$ and $\sigma(M,z)$ is the rms of the linear density
field smoothed by a top-hat in $R$ that encloses $M$ at the background density. Specifically,
the mass function takes the familiar form 
\be
 n_{\ln M}(M,z) = \frac{\bar \rho_m}{M}\frac{d\nu}{d\ln M} f(\nu) \,,
\ee
and the Lagrangian bias is given by
\be
 b^L_1(M,z) \equiv \frac{\partial \ln n_{\ln M}}{\partial\delta_m}
 = \frac{d \ln n_{\ln M}}{d\nu}\frac{1}{\sigma(M,z)}\frac{d\delta_{c}}{d\delta_m} \,,
\label{eq:bLMari}
\ee
where ${d\delta_{c }}/{d\delta_m}$ is the response of the collapse threshold {\it at}
$z$ to the {\it entire} growth history of a long-wavelength mode which reaches $\delta_m$
at $z$ in a manner that depends on $k$. The partial derivative denotes the fact that
it is at fixed comoving volume and does not account for the dilation induced by $\delta_m$.
In this way, the model differs fundamentally from ones that assume halo bias is local
in the density field at a single epoch. On the other hand the association of the observed
redshift $z$ with the collapse or formation redshift of all halos is ad hoc since real
halos continually grow in mass and merge with each other. Predictions would differ
substantially if for example this redshift were instead associated with the last major
merger of a halo.

The collapse threshold $\delta_{c}(z)$ is computed by numerically solving for the evolution
of a top-hat density perturbation of size $R$ enclosing mass $M = 4\pi R^3{\bar\rho_m}/3$
in the presence of the long-wavelength mode and its whole evolutionary history, $\delta_m(z)$.
If the components other than CDM (here quintessence) do not cluster on the scale of the
top-hat, the evolution of the radius $R$ is given by
\ba
 \ddot{R}=\:&-\frac{4\pi G}{3} \sum_{J\neq \rm CDM}
 \left[ \bar\rho_J(t)+ 3\bar p_J(t) + \delta\rho_J(t) + 3\delta p_J(t)\right] R \vs
 \:&-\frac{GM(<R)}{R^2} \,,
\label{eq:ddotRLW}
\ea
where $\delta\rho_J$ and $\delta p_J$ are the long-wavelength perturbations in the non-CDM
components that may exist along with $\delta_m(z)$. \refEq{ddotRLW} is solved with the
initial conditions 
\ba
 R_i \:&= \frac{3M}{(4\pi {\bar\rho_m})^{1/3}}\left[ 1-\frac{\delta_i + \delta_{m}(z_i) }{3}\right] \,, \vs
 \dot{R}_i\:&= H_i R_i\left[ 1-\frac{ \dot{\delta}_i +\dot \delta_{m}(z_i) }{3H_i }\right] \,,
\ea
where $\delta_i$ is the initial amplitude of the perturbation on scale $R$ and $\delta_m(z_i)$
is the initial amplitude of the long-wavelength mode. The collapse threshold is the critical
value of $\delta_i$ needed to collapse ($R\rightarrow 0$) at $z$ linearly extrapolated to $z$
\be
 \delta_{c}(z) = \frac{D^{\downarrow}(z)}{D^{\downarrow}(z_i)}\delta_i \,.
\ee
Here we have assumed that the top-hat radius is always below the Jeans scale. The response
of the collapse threshold to the long-wavelength mode is
\be
 \frac{d\delta_{c}}{d\delta_m}(z) = \frac{\delta_{c}(z |\delta_m) - \delta_{c}(z)}{\delta_m(z)} \,.
\ee
For further details see \cite{LoVerde:2014pxa, LoVerde:2014rxa}.

Note that for $z_i$ deep in the matter dominated era the linear growth rates of $\delta_i$
and $\delta_m(z_i)$ are the same both above and below the Jeans scale. Below the Jeans
scale, the linear evolution of the long-wavelength mode is  entirely the same as the
linear evolution of the mode on scale $R$, and we recover the usual result ${d\delta_{c}}/{d\delta_m} =-1$
that makes bias scale-free below the Jeans scale at the observation epoch.

Above the Jeans scale, this relation is no longer true since the long-wavelength mode
grows faster than the linear evolution of the mode on scale $R$. The prediction for
the difference between the Lagrangian bias factors on sub-Jeans ($\downarrow$) and
super-Jeans ($\uparrow$) Jeans scales is then
\be
\label{eq:L14}
 \frac{b^{L\uparrow}_1 - b^{L\downarrow}_1}{b^{L\downarrow}_1}
 =-\left(\frac{d\delta_{c}}{d\delta_{\uparrow } }+ 1\right) \,.
\ee
In \reffig{bLversusmodels}, we compare this prediction to the separate universe response
bias at $z=0$ and $z=1$. These predictions better capture the temporal nonlocality of halo
bias than the PHS11 model. Had we assumed, contrary to fact, that the change in the collapse threshold is determined
at the initial epoch $\delta_i(\delta_m) = \delta_i - \delta_m(z_i)$ then
$d\delta_c/d\delta_m(z)= -D^\downarrow(z)/D^\uparrow(z)$ and we would recover the PHS11 prediction.
Instead the L14 model assumes  a universal mass function
in the peak height $\nu$ of the spherical collapse threshold at the observation epoch.

\bibliography{fsuq}

\end{document}